\documentclass[twoside,titlepage,final,12pt]{article}
\usepackage{latexsym}
\usepackage{amssymb}

\usepackage{graphicx}
\usepackage[pdftex,dvipsnames,usenames]{color}
\definecolor{IITred}{rgb}{0.5,0.05,0.05}

\usepackage[colorlinks,citecolor=MidnightBlue,urlcolor=Sepia,linkcolor=ForestGreen]{hyperref}

\catcode`\@=11
\def\ps@ppt{\def\@oddhead{\qquad LHC 
Physics Potential---2011 Run\hfil \thepage\qquad}\def\@evenhead{\qquad\thepage \hfil {Chris Quigg} \qquad}
\def\@oddfoot{}\def\@evenfoot{}}    
\catcode`\@=12

\def\urll#1#2{\mbox{\href{#1}{\sf #2}}}

\makeatletter
   \renewcommand{\section}{\@startsection{section}{1}{0mm}
   {\baselineskip}%
   {\baselineskip}{\normalfont\normalsize\centering}}%
   \makeatother

\newcommand{\eqn}[1]{(\ref{#1})}

\newcommand{\gev}{\ensuremath{\hbox{ GeV}}}
\newcommand{\tev}{\ensuremath{\hbox{ TeV}}}
\newcommand{\pb}{\ensuremath{\hbox{ pb}}}
\newcommand{\fb}{\ensuremath{\hbox{ fb}}}

\def\phystoday#1#2#3#4{\frenchspacing{\it Phys. Today }{\bf #1}, #2 (\ifcase#3\or January\or 
         February\or March\or April\or May\or June\or July\or August\or 
         September\or October\or November\or December\fi, 19#4)}
\def\slashii#1{\setbox0=\hbox{$#1$}             
   \dimen0=\wd0                                 
   \setbox1=\hbox{\sl/} \dimen1=\wd1            
   \ifdim\dimen0>\dimen1                        
      \rlap{\hbox to \dimen0{\hfil\sl/\hfil}}   
      #1                                        
   \else                                        
      \rlap{\hbox to \dimen1{\hfil$#1$\hfil}}   
      \hbox{\sl/}                               
   \fi}                                         %

\begin{document}
\begin{flushright}
	FERMILAB--FN--0913--T \\ Rev. February 1, 2011
\end{flushright}
\vspace*{\stretch{1}}
\begin{center}
	{\Huge  LHC Physics Potential \textit{vs.}~Energy: \\ Considerations for the 2011 Run} 
	\\[5mm]
	{\large Chris Quigg* } \\[5mm]
	Theoretical Physics Department \\
	Fermi National Accelerator Laboratory\\ Batavia, Illinois 60510 USA \\[2mm] and \\[2mm]
	CERN, Department of Physics, Theory Unit \\ CHÐ1211 Geneva 23, Switzerland \\[8mm]
	\parbox{4.in}{Parton luminosities are convenient for estimating how the physics potential of Large Hadron Collider experiments depends on the energy of the proton beams.  I quantify the advantage of increasing the beam energy from $3.5\tev$ to $4\tev$. I present parton luminosities, ratios of parton luminosities, and contours of fixed parton luminosity for $gg$, $u\bar{d}$, $qq$, and $gq$ interactions over the energy range relevant to the Large Hadron Collider, along with example analyses for specific processes. This note extends the analysis presented in Ref.~\cite{Quigg:2009gg}. Full-size figures are available as pdf files at \urll{http://lutece.fnal.gov/PartonLum11/}{lutece.fnal.gov/PartonLum11/}.}	
\end{center}
\vspace*{\stretch{1.5}}
		*\urll{mailto:quigg@fnal.gov}{E-mail:quigg@fnal.gov}
\newpage
\setlength{\parindent}{2ex}
\setlength{\parskip}{12pt}
\section{Preliminaries}
\noindent
The 2009-2010 run of the Large Hadron Collider at CERN is complete, with the delivery of some $45\pb^{-1}$ of proton-proton collisions at $3.5\tev$ per beam to the ATLAS and CMS experiments. The primary objective of the run, to commission and ensure stable operation of the accelerator complex and the experiments, has been achieved, and much has been learned about machine operation. The experiments succeeded in ``rediscovering'' the standard model of particle physics, and using familiar physics objects such as $W^\pm$, $Z^0$, $J\!/\!\psi$, $\Upsilon$, jets, $b$-hadrons, and top-quark pairs to tune detector performance. In a few cases, LHC experiments have begun to explore virgin territory and surpass the discovery reach of the Tevatron experiments CDF and D0~\cite{LHCjamboree}.

Coming soon is an extended physics run during 2011-2012, with the goal of logging $1\fb^{-1}$ by the end of 2011 and perhaps $5\fb^{-1}$ by the end of 2012. Still to be decided is whether to raise the proton energy from $3.5\tev$ to $4\tev$ per beam. The object of this note is to quantify the enhanced sensitivity that would follow from running at $\sqrt{s}=8\tev$ rather than $\sqrt{s}=7\tev$ by considering some key parton luminosities. The choice of energy impacts the comparison of sensitivities at the LHC and the Tevatron. Taking the long view, it is also important to keep in mind how much is to be gained by approaching the LHC design energy of $\sqrt{s}=14\tev$.

Detailed simulations of signals and backgrounds are of unquestioned value for in-depth consideration of the physics possibilities, especially in view of experience gained during the 2009-2010 run of the LHC. Higher-order perturbative-QCD calculations add insight~\cite{Alekhin:2010dd}. However, much can be learned about the general issues of energy, luminosity, and the relative merits of proton-proton and proton-antiproton collisions by comparing the luminosities of parton-parton collisions as a function of $\sqrt{\hat{s}}$, the c.m. energy of the colliding partons~\cite{Eichten:1984eu}. [A high-energy proton is, in essence, a broadband unseparated beam of quarks, antiquarks, and gluons.] By contemplating the parton luminosities in light of existing theoretical and experimental knowledge, physicists should be able to anticipate and critically examine the broad results of Monte Carlo studies. The more prior knowledge a user brings to the parton luminosities, the more useful insights they can reveal.

Taking into account the $1/\hat{s}$ behavior of the hard-scattering processes that define much of the physics motivation for a multi-TeV hadron collider, the \textit{parton luminosity} 
\begin{equation}
\frac{\tau}{\hat{s}}\frac{d\mathcal{L}_{ij}}{d\tau}  \equiv  \frac{\tau/\hat{s}}{1 + \delta_{ij}}\int_\tau^1 \!\!dx[f_i^{(a)}(x) f^{(b)}_j(\tau/x)  + f_j^{(a)}(x)f_i^{(b)}(\tau/x)]/x, 
 \label{eq:lumdef}
 \end{equation}
 which has dimensions of a cross section, is a convenient measure of the reach of a collider of given energy and hadron-hadron luminosity. Here $f_i^{(a)}(x)$ is the number distribution of partons of species $i$ carrying momentum fraction $x$ of hadron $a$. For hadrons colliding with c.m. energy $\sqrt{s}$, the scaling variable $\tau$ is given by
 \begin{equation} 
 \tau  = \hat{s}/s .
 \label{eq:taudef}
 \end{equation}
 
The  cross section for the hadronic reaction
 \begin{equation}
 a + b \to \alpha + \hbox{anything}
 \label{eq:abcX}
 \end{equation}
 is given by 
 \begin{equation}
\sigma(s) = \sum_{\{ij\}}\int_{\tau_0}^1 \frac{d\tau}{\tau}\cdot{\frac{\tau}{\hat{s}}\frac{d\mathcal{L}_{ij}}{d\tau}}\cdot\left[\hat{s}\hat{\sigma}_{ij \to \alpha}(\hat{s})\right] ,
\label{eq:cross}
\end{equation}
 where $\hat{\sigma}_{ij \to \alpha}$ is the operative parton-level cross section. The (dimensionless) factor in square brackets is approximately determined by couplings. Many explicit (leading-order) forms of $\hat{\sigma}$ are given in Refs.~\cite{Eichten:1984eu}. The logarithmic integral typically gives a factor of order unity.

 If event rates for signal and backgrounds are known---by calculation or by measurement---for some point $(\sqrt{s},\sqrt{\hat{s}})$, the parton luminosities can be used to estimate the rates at other points, at an accuracy satisfactory for orientation. Because leading-order parton distributions have a simple intuitive interpretation, I have chosen the CTEQ6L1 leading-order parton distributions~\cite{Pumplin:2002vw} for the calculations presented in this note. Any of the modern sets of parton distributions will yield similar results for energy-to-energy comparisons~\cite{Stirling}.

\begin{figure}[tb]
\centerline{\includegraphics[height=0.7\textheight]{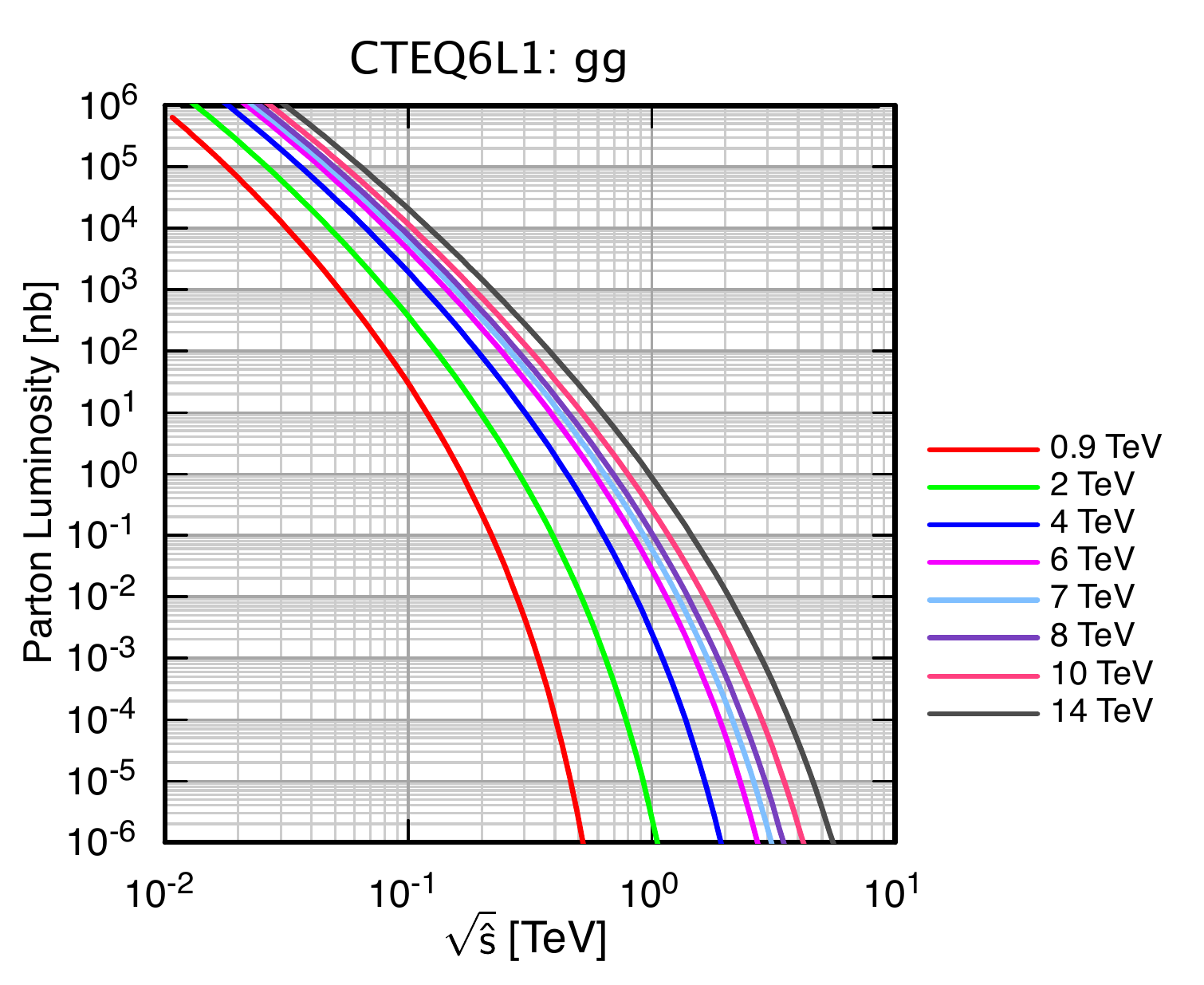}}
\caption{Parton luminosity $(\tau/\hat{s})d\mathcal{L}/d\tau$ for $gg$ interactions.}
\label{fig:gg2}
\end{figure}

\begin{figure}[p]
\centerline{\includegraphics[height=0.7\textheight]{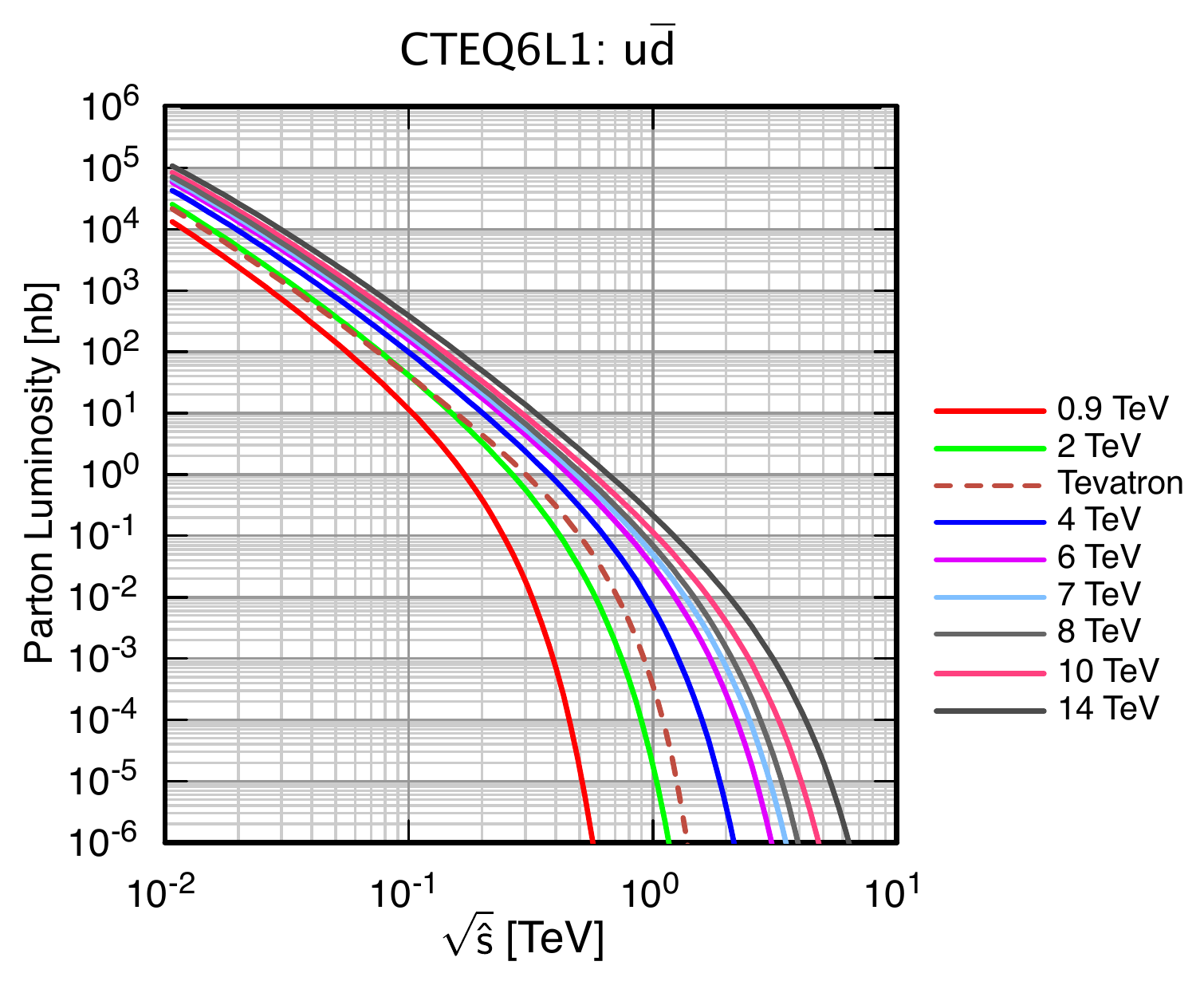}}
\caption{Parton luminosity $(\tau/\hat{s})d\mathcal{L}/d\tau$ for $u\bar{d}$ interactions.}
\label{fig:udbar2}
\end{figure}

\begin{figure}[p]
\centerline{\includegraphics[height=0.7\textheight]{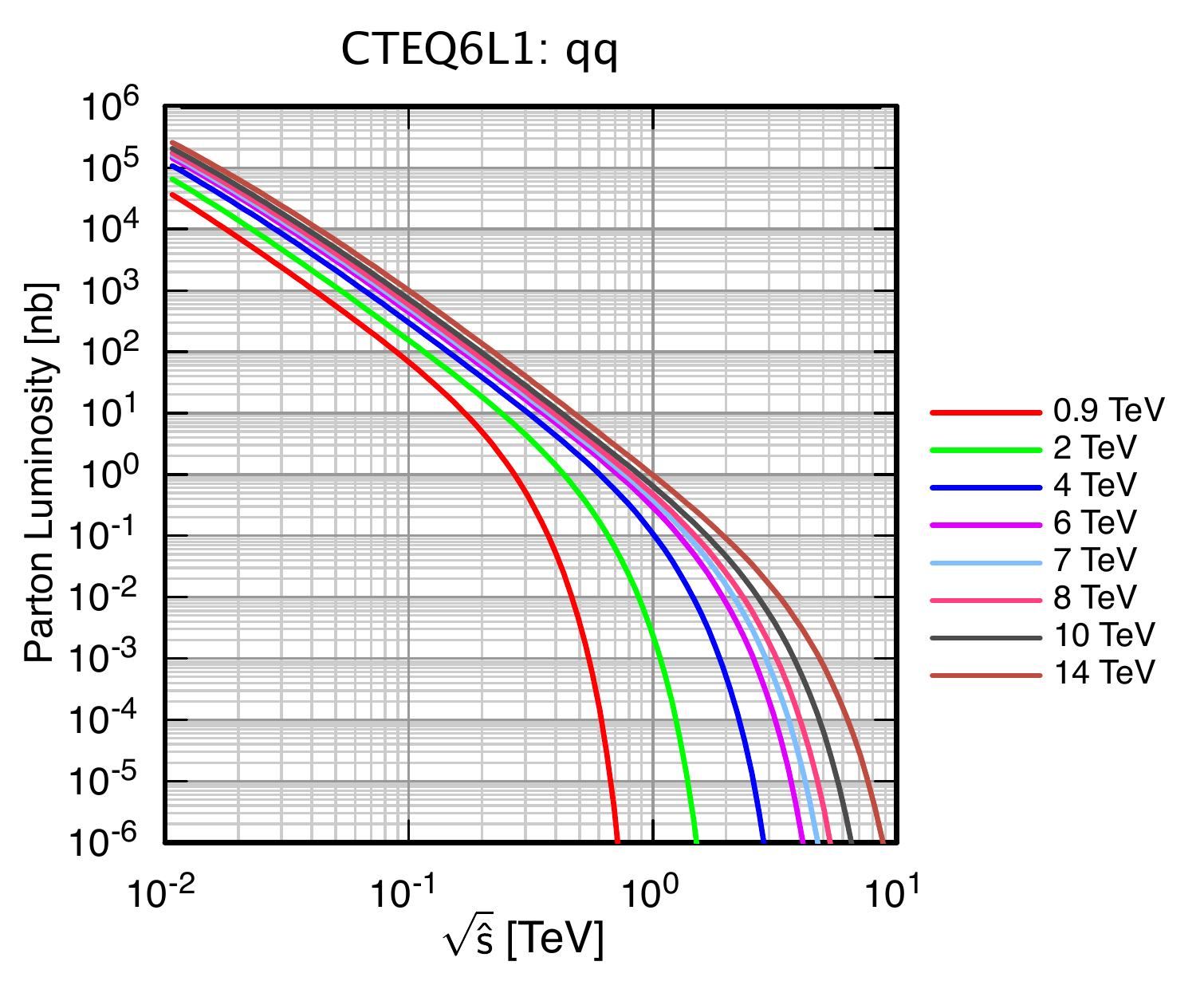}}
\caption{Parton luminosity $(\tau/\hat{s})d\mathcal{L}/d\tau$ for $qq$ interactions. See the text surrounding \eqn{eq:qqdef} and \eqn{eq:qqdefbar} for definitions.}
\label{fig:qq2}
\end{figure}

\begin{figure}[p]
\centerline{\includegraphics[height=0.7\textheight]{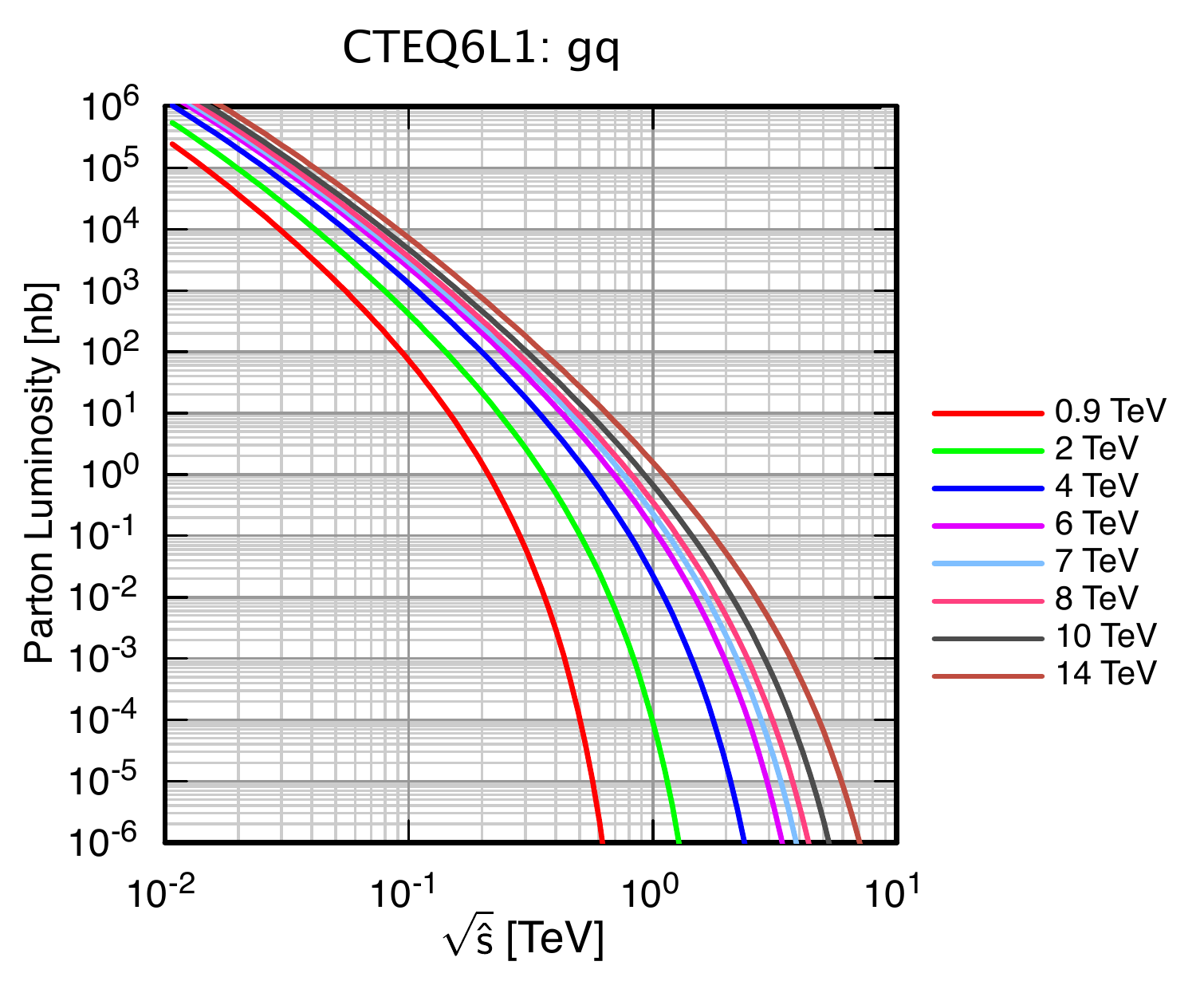}}
\caption{Parton luminosity $(\tau/\hat{s})d\mathcal{L}/d\tau$ for $gq$ interactions. See the text surrounding \eqn{eq:gqdef} for definitions.}
\label{fig:gq2}
\end{figure}

I present four examples relevant to discovery physics: gluon-gluon interactions, $u\bar{d}$ interactions, interactions among generic light quarks, and interactions of gluons with light quarks or antiquarks. The parton luminosities for gluon-gluon interactions are given in Figure~\ref{fig:gg2}. These are identical for $pp$ and $\bar{p}p$ collisions. The parton luminosities for $u\bar{d}$ interactions are plotted in Figure~\ref{fig:udbar2}. In $pp$ collisions, $u\bar{d}$ is a valence--sea combination; in $\bar{p}p$ collisions, it is valence--valence. The difference is reflected in the excess of the Tevatron luminosities in Figure~\ref{fig:udbar2} over the proton-proton luminosities at $\sqrt{s}= 2\tev$.
The parton luminosities for light-quark--light-quark interactions in $pp$ collisions are displayed in Figure~\ref{fig:qq2}, as examples of valence--valence interactions leading to final states such as two jets. What is plotted here is the combination
\begin{equation}
(u + d)^{(1)}\otimes (u + d)^{(2)} .
\label{eq:qqdef}
\end{equation}
For $\bar{p}p$ collisions at the Tevatron, interpret the 2-TeV curve as
\begin{equation}
(u + d)^{(p)}\otimes (\bar{u} + \bar{d})^{(\bar{p})} ;
\label{eq:qqdefbar}
\end{equation}
these valence--valence interactions are the main source of high-transverse-momentum jets at the Tevatron. The parton luminosities for gluon--light-quark interactions are shown in Figure~\ref{fig:gq2}. The quantity displayed is the combination
\begin{equation}
(u + d)^{(1)} \otimes g^{(2)} .
\label{eq:gqdef}
\end{equation}
In $pp$ collisions, the $gq$ luminosity is twice what is shown in Figure~\ref{fig:gq2}, corresponding to $(u + d)^{(1)} \otimes g^{(2)} + g^{(1)} \otimes (u + d)^{(2)}$. For $\bar{p}p$ scattering at the Tevatron, interpret the 2-TeV curve as either $(u + d)^{(p)} \otimes g^{(\bar{p})}$ ($gq$ collisions) or $g^{(p)} \otimes(\bar{u} + \bar{d})^{(\bar{p})}$ ($g\bar{q}$ collisions).

\textit{Ratios} of parton luminosities are especially useful for addressing what is gained or lost by running at one energy instead of another. Let us consider each of the example cases in turn.

\section{Gluon-gluon interactions}
\begin{figure}[p]
\centerline{\includegraphics[height=0.7\textheight]{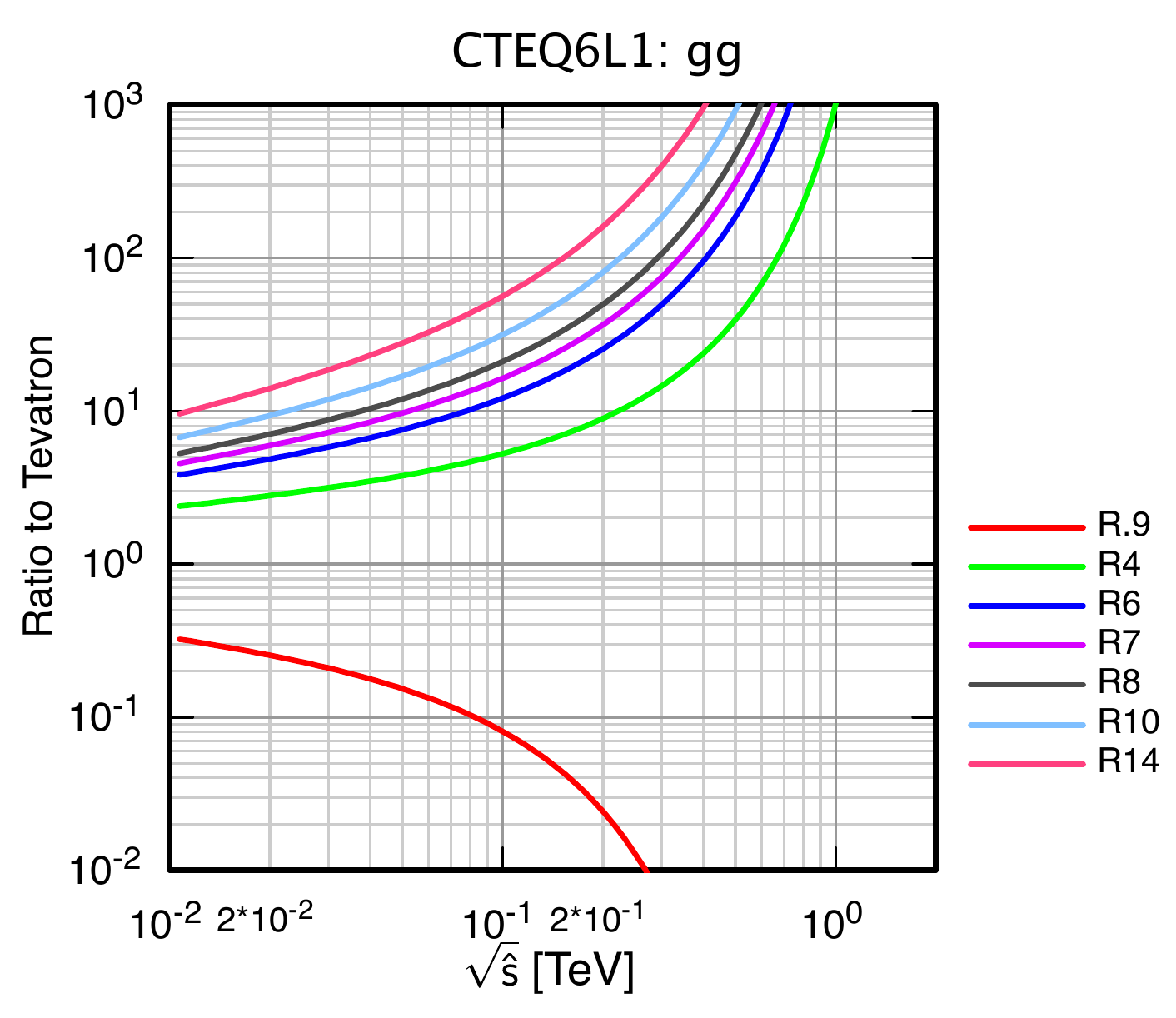}}
\caption{Comparison of parton luminosity for $gg$ interactions at specified energies with luminosity at $2\tev$.}
\label{fig:ggrat2}
\end{figure}

\begin{figure}[p]
\centerline{\includegraphics[height=0.7\textheight]{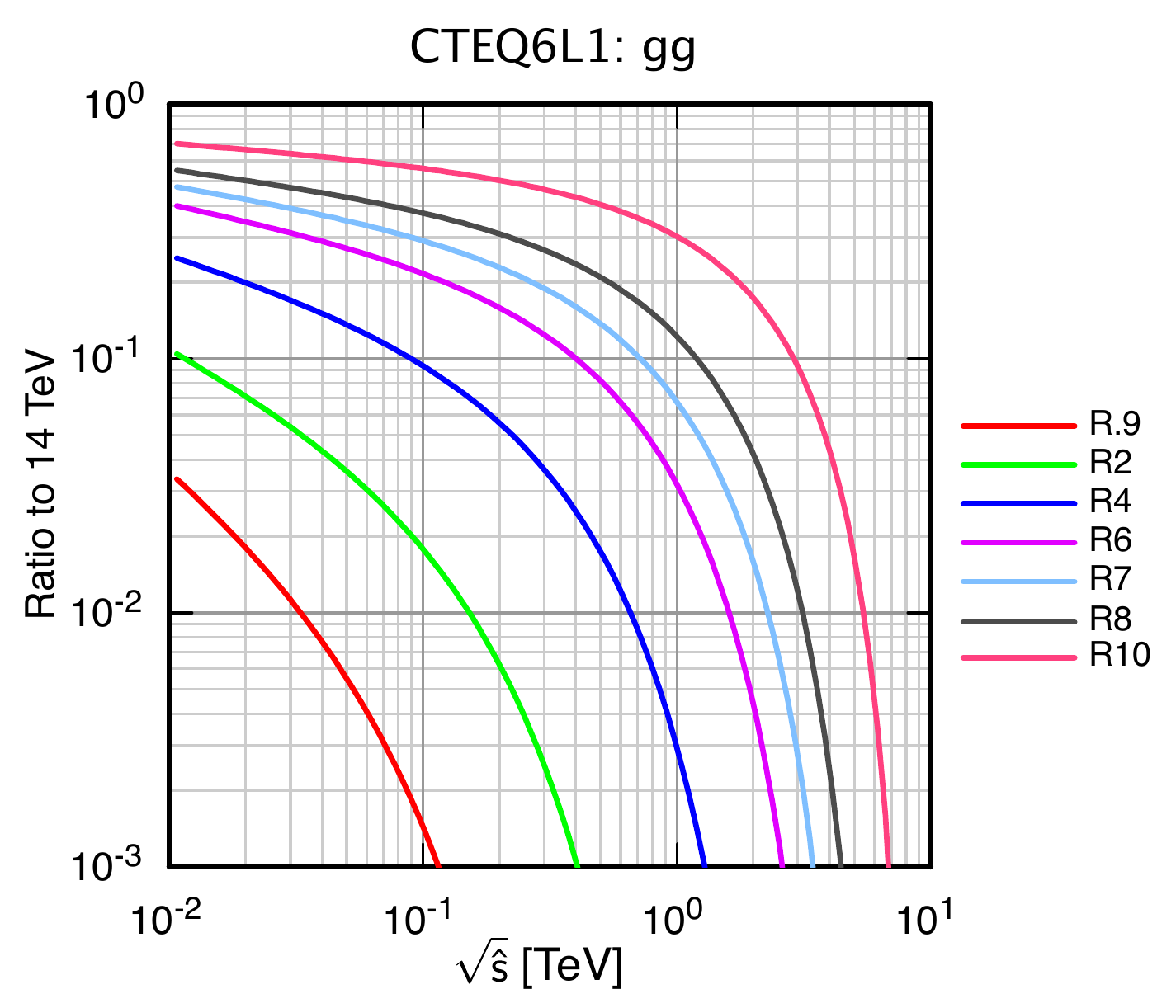}}
\caption{Comparison of parton luminosity for $gg$ interactions at specified energies with luminosity at $14\tev$.}
\label{fig:ggrat14}
\end{figure}

Ratios of parton luminosities for gluon-gluon interactions in $p^\pm p$ collisions at specified energies to the $gg$ luminosity at the Tevatron are shown in Figure~\ref{fig:ggrat2}; ratios to the LHC at design energy in Figure~\ref{fig:ggrat14}. At $\sqrt{\hat{s}} \approx 0.4\tev$, characteristic of $t\bar{t}$ pair production, Figure~\ref{fig:ggrat2} shows that the $gg$ luminosity rises by three orders of magnitude from the 2-TeV Tevatron to the 14-TeV LHC. This rise is the source of the computed increase in the $gg \to t\bar{t}$ cross section from Tevatron to LHC, and is the basis for the (oversimplified) slogan,  ``The Tevatron is a quark collider, the LHC is a gluon collider." Figure~\ref{fig:ggrat14} shows that the $gg \to t\bar{t}$ yield drops by a bit more than a factor of 6 between $14\tev$ and $7\tev$. To first approximation, accumulating a $t\bar{t}$ sample of specified size at $\sqrt{s}=7\tev$ will require about $6\times$ the integrated luminosity that would have been needed at $\sqrt{s}=14\tev$, although acceptance cuts should have less effect at the lower energy. At $\sqrt{s} = 10\tev$, the $gg \to t\bar{t}$ rate is a factor of $2.3$ smaller than at design energy. The $gg$ luminosity at $\sqrt{\hat{s}} = 0.4\tev$ is approximately $1.47\times$ higher at $\sqrt{s}=8\tev$ than at $7\tev$.

\begin{figure}[p]
\centerline{\includegraphics[height=0.7\textheight]{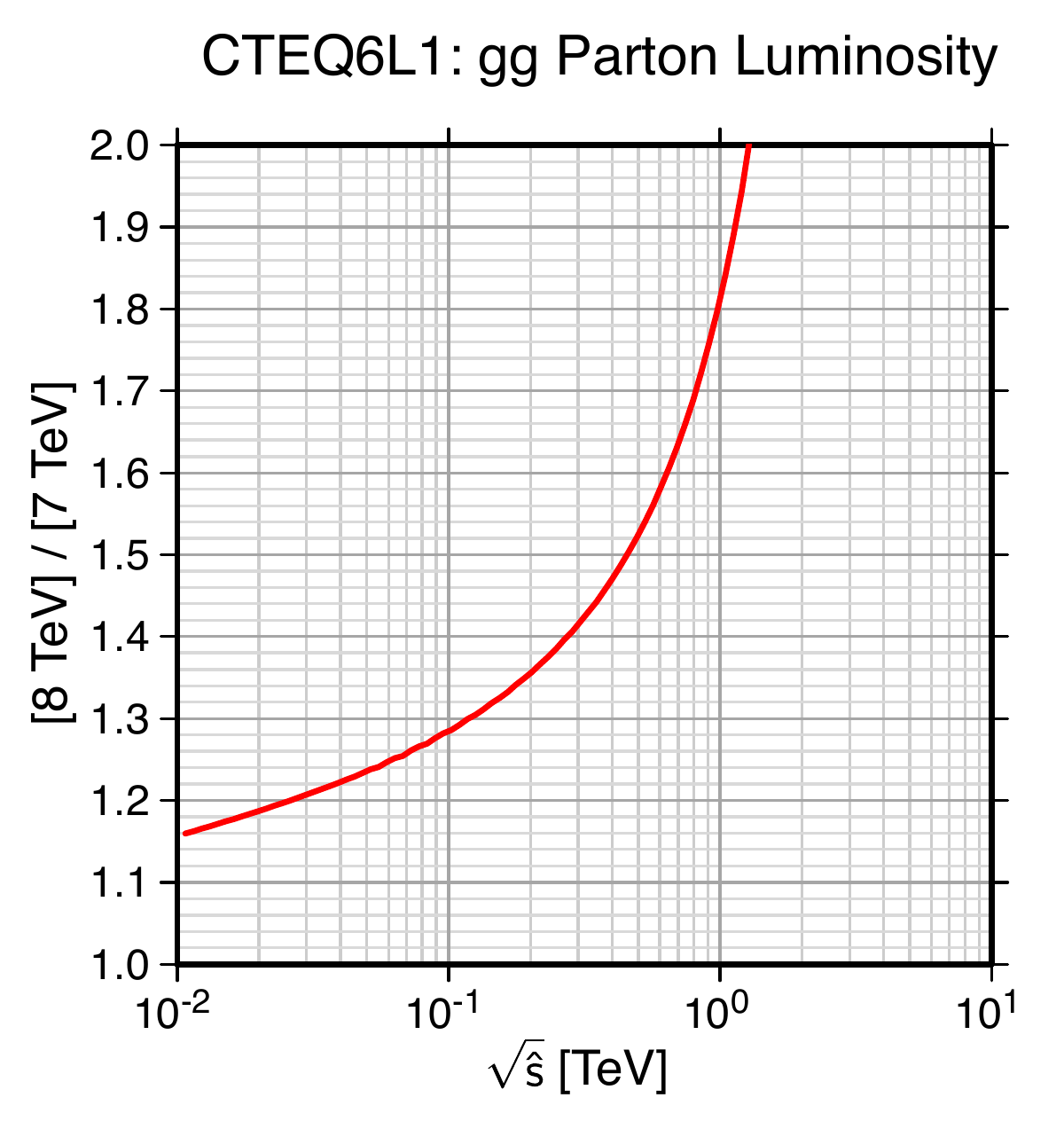}}
\caption{Ratio of parton luminosity for $gg$ interactions in $pp$ collisions at $\sqrt{s} = 8\tev$ to luminosity at $7\tev$.}
\label{fig:ggrat87}
\end{figure}

\begin{figure}[p]
\centerline{\includegraphics[height=0.7\textheight]{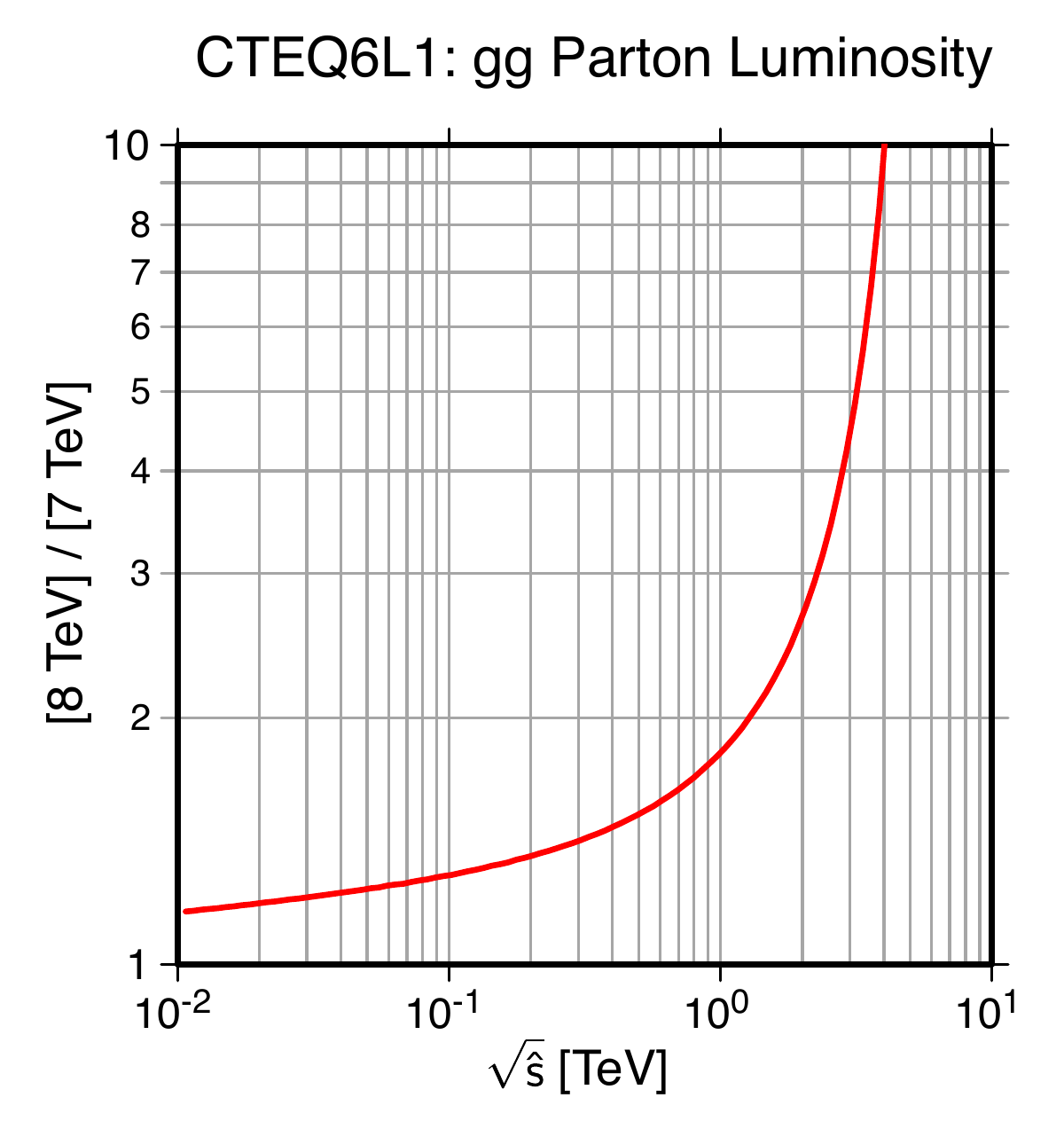}}
\caption{Ratio of parton luminosity for $gg$ interactions in $pp$ collisions at $\sqrt{s} = 8\tev$ to luminosity at $7\tev$ (logarithmic scale).}
\label{fig:ggrat87log}
\end{figure}

The dominant mechanism for light Higgs-boson production at both the Tevatron and the LHC is $gg \to \hbox{top-quark loop} \to H$, so the rates are controlled by the $gg$ luminosity. For $M_H \approx 120\gev$, the $gg$ luminosity is approximately $(20, 25, 38, 70)\times$ larger at $\sqrt{s} = (7, 8, 10, 14)\tev$ than at the Tevatron. LHC experiments are likely to rely on the rare $\gamma\gamma$ decay of a light Higgs boson, for which high integrated luminosities will be required. At somewhat higher Higgs-boson masses, the situation could be more promising for early running. For $M_H = 175\gev$, a mass at which $H \to ZZ$ becomes a significant decay mode, the $gg$ luminosity is roughly $(30, 40, 65, 130)\times$ larger at $\sqrt{s} = (7, 8, 10, 14)\tev$ than at the Tevatron. The potential Tevatron sensitivity for $gg \to H \to ZZ$, based on the current integrated luminosity of $10\fb^{-1}$ would be matched at the LHC by integrated luminosities of $(340, 250, 160, 80)\pb^{-1}$ at $\sqrt{s} = (7, 8, 10, 14)\tev$. Note that these levels do not correspond to the thresholds needed for discovery (although those could be worked out, given a discovery criterion), but to the point at which the LHC would begin to break new ground, compared to the Tevatron sample now in hand. The parton-luminosity advantage of 8-TeV over 7-TeV running is $1.3$ for 120-GeV $gg$ collisions, $1.34$ for 175-GeV $gg$ collisions, and $1.47$ for 400-GeV $gg$ collisions (see Figure~\ref{fig:ggrat87}). By $\sqrt{\hat{s}} \approx 4\tev$, the multiplier reaches an order of magnitude (see Figure~\ref{fig:ggrat87log}).

\section{$u\bar{d}$ interactions}
\begin{figure}[p]
\centerline{\includegraphics[height=0.7\textheight]{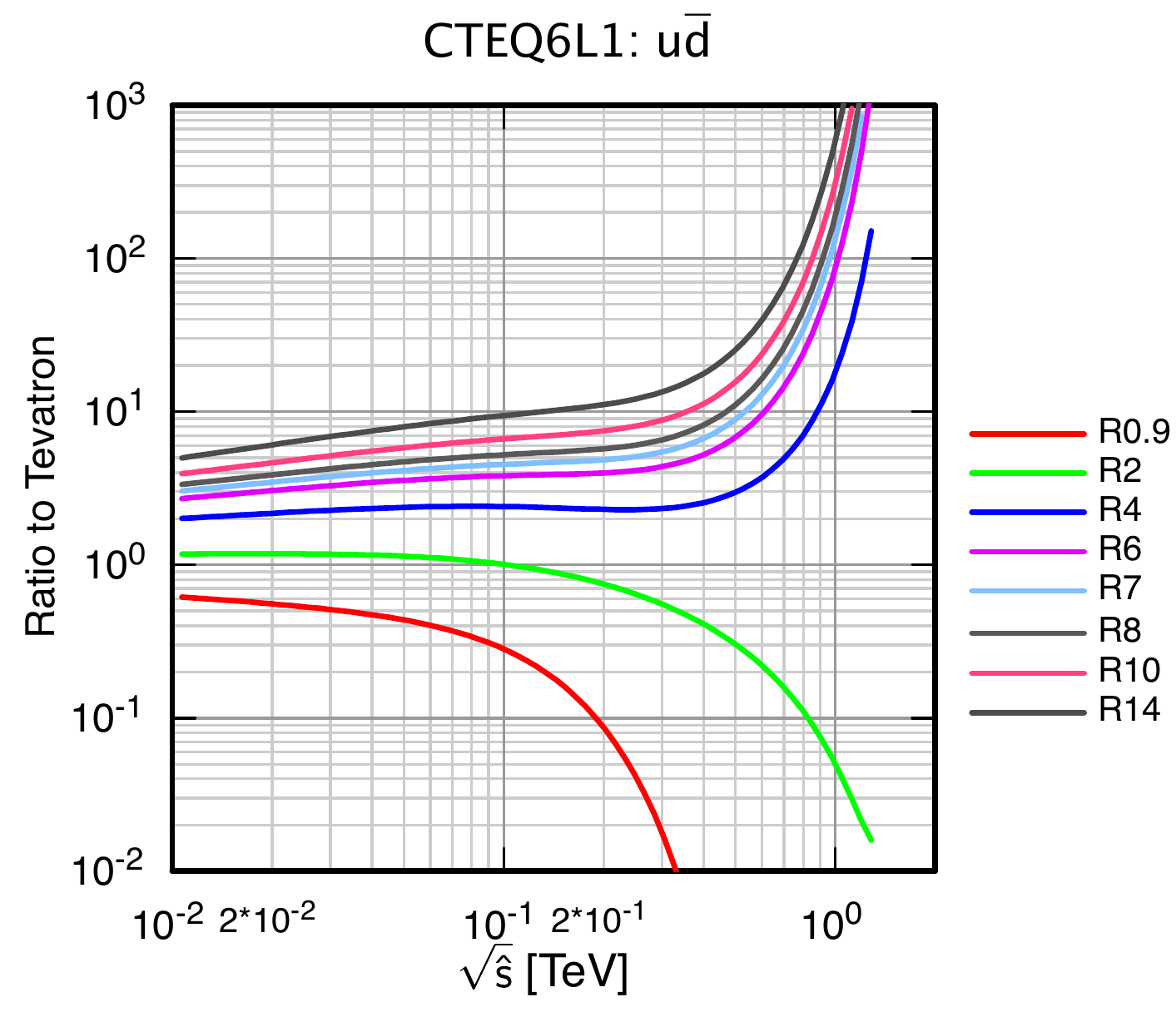}}
\caption{Comparison of parton luminosity for $u\bar{d}$ interactions in $pp$ collisions at specified energies with luminosity in $\bar{p}p$ collisions at $2\tev$.}
\label{fig:udbarrat2}
\end{figure}

\begin{figure}[p]
\centerline{\includegraphics[height=0.7\textheight]{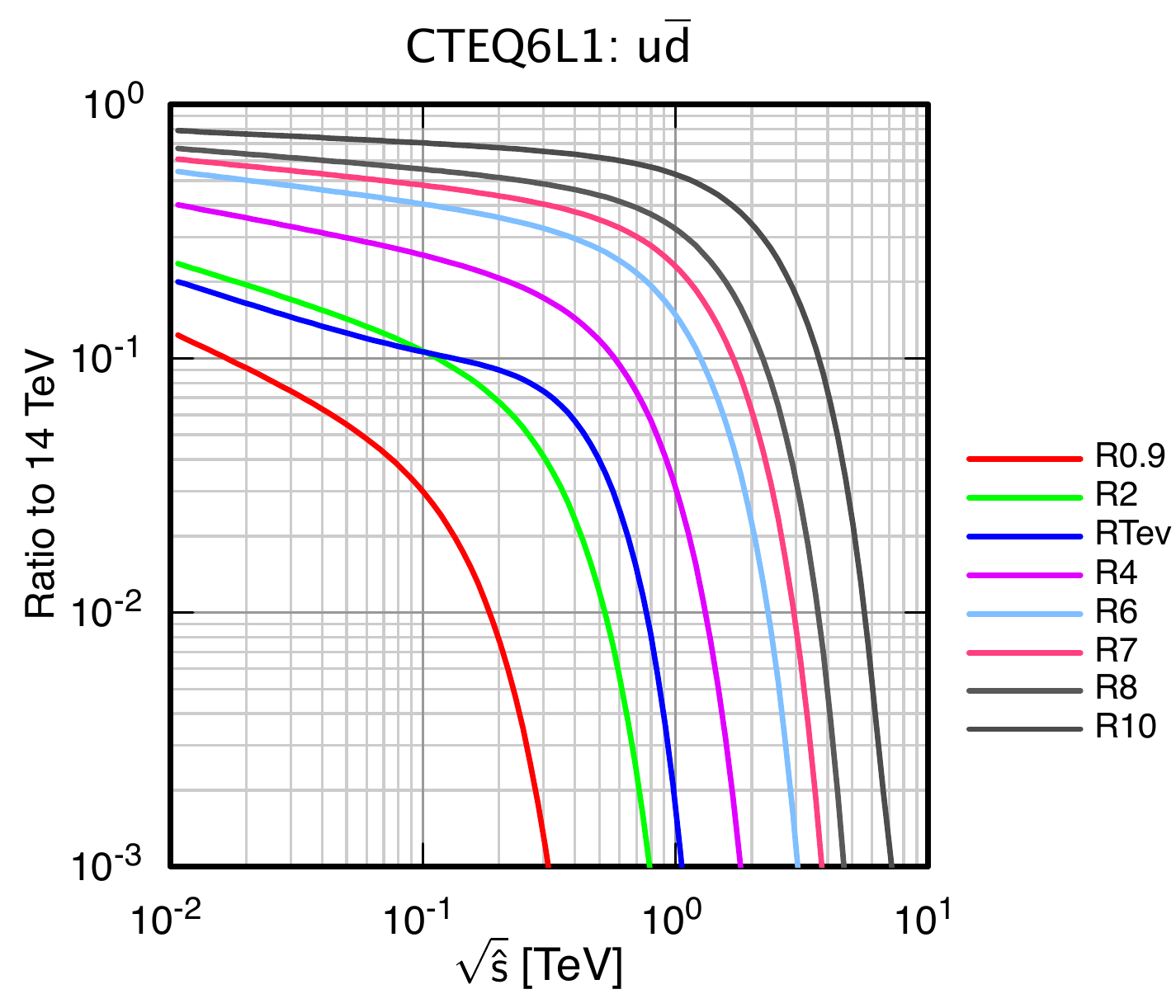}}
\caption{Comparison of parton luminosity for $u\bar{d}$ interactions at specified energies with luminosity at $14\tev$.}
\label{fig:udbarrat14}
\end{figure}

Ratios of parton luminosities for $u\bar{d}$ interactions in $pp$ collisions at specified energies to the $u\bar{d}$ luminosity in $\bar{p}p$ collisions at the Tevatron are plotted in Figure~\ref{fig:udbarrat2}. Ratios to the LHC at design energy are shown in Figure~\ref{fig:udbarrat14}. These ratios of luminosities apply directly to the production of $W$ bosons and to the search for new $W^\prime$ bosons. They are also indicative of the behavior of $u\bar{u}$ and $d\bar{d}$ luminosities, which enter the production of $Z$, $Z^\prime$, and $W^+W^-$ or $ZZ$ pairs that are backgrounds to Higgs-boson searches for $M_H \gtrsim 140\gev$. For the case of $W^+$ production, Figure~\ref{fig:udbarrat2} shows that the rates will be higher by factors of $(4.4, 5, 6.4, 9)$ at $\sqrt{s} = (7, 8, 10, 14)\tev$, compared to the Tevatron rate. At an invariant mass of $175\gev$, the curves in Figure~\ref{fig:udbarrat2} show that the rate for the background processes $q\bar{q} \to VV$ ($V = W,Z$) grows less rapidly than the rate for the signal process $gg \to H$ discussed in the previous paragraph. The enhancements over the Tevatron are by factors of $(4.8,5.6, 7.3,10.7)$ at $\sqrt{s} = (7, 8, 10, 14)\tev$. 

For a $W^\prime$ search at $M_{W^\prime} = 1\tev$, the production rates are larger by factors of $(134, 190, 308, 580)$ at $\sqrt{s} = (7, 8, 10, 14)\tev$, so the Tevatron's $10\fb^{-1}$ sensitivity would be matched at integrated luminosities of approximately $(75, 53, 33, 17)\pb^{-1}$, before taking into account relative detector acceptances. At still higher masses, the penalty for LHC running below design energy is correspondingly greater. At $M_{W^\prime} = 2\tev$, the rates are diminished by factors of approximately $2.95, 7.8$, and $16$ at $\sqrt{s}=(10, 8, 7)\tev$. For high-mass searches, one must refer back to the parton luminosities themselves (Figure~\ref{fig:udbar2}) to check whether the absolute rates give adequate sensitivity.

\begin{figure}[p]
\centerline{\includegraphics[height=0.7\textheight]{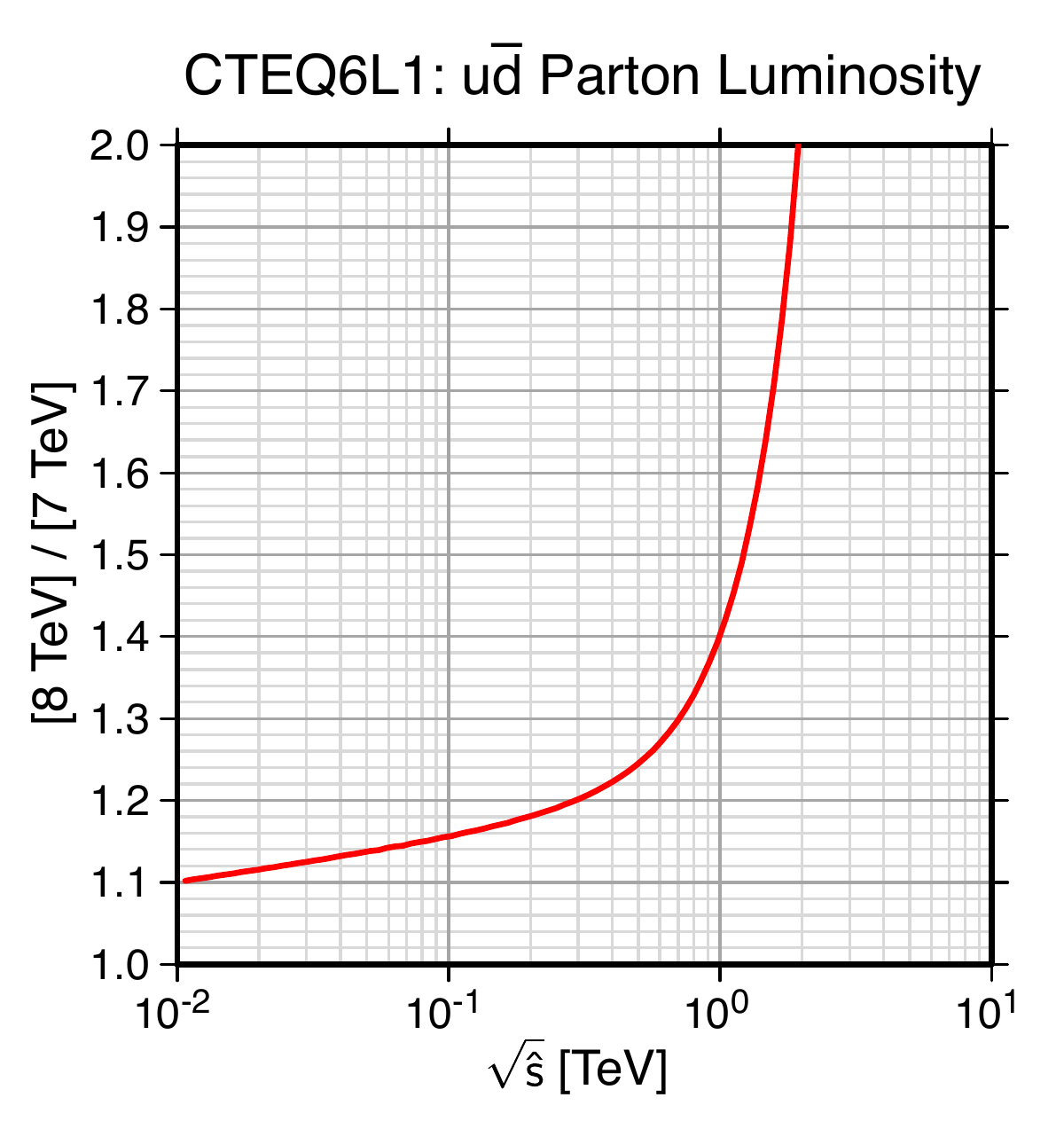}}
\caption{Ratio of parton luminosity for $u\bar{d}$ interactions in $pp$ collisions at $\sqrt{s} = 8\tev$ to luminosity at $7\tev$.}
\label{fig:udbar87}
\end{figure}

\begin{figure}[p]
\centerline{\includegraphics[height=0.7\textheight]{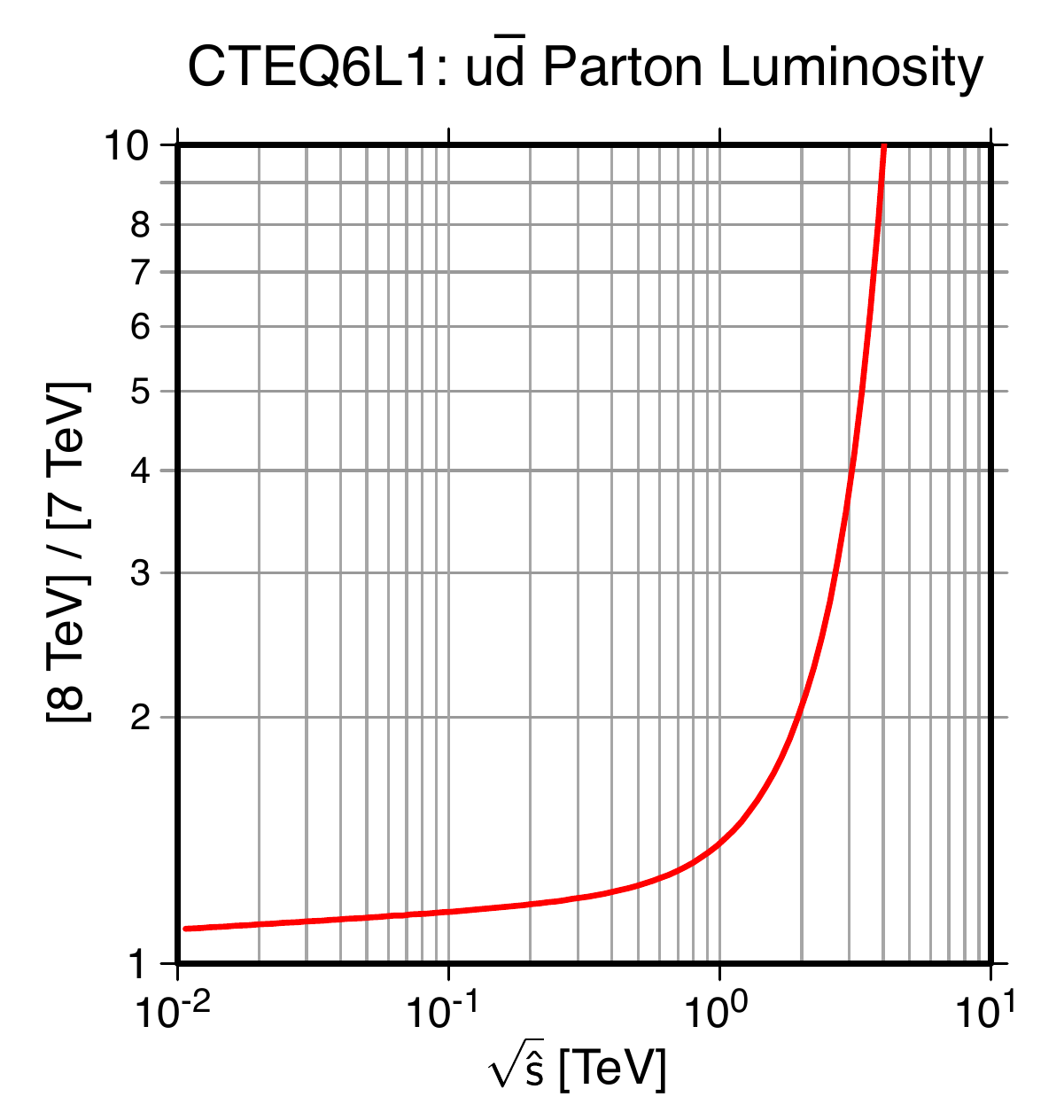}}
\caption{Ratio of parton luminosity for $u\bar{d}$ interactions in $pp$ collisions at $\sqrt{s} = 8\tev$ to luminosity at $7\tev$ (logarithmic scale).}
\label{fig:udbar87log}
\end{figure}

The $q\bar{q}$ contribution to $t\bar{t}$ production will also track the ratios of $u\bar{d}$ luminosities. In the range of interest, $\sqrt{\hat{s}} \approx 0.4\tev$, the rate is enhanced over the Tevatron ($\bar{p}p$ collisions!) rate by factors of roughly $(7, 8, 11, 18)$ at $\sqrt{s} = (7, 8, 10, 14)\tev$, far smaller than the enhancements we noted above for the $gg \to t\bar{t}$ rates. The behavior of the $u\bar{d}$ parton luminosities also determines how rates for $H(W^\pm,Z)$ associated production and for pair production of new colored particles (e.g., superpartners) scale under different operating conditions. 
Figure~\ref{fig:udbar87} summarizes the $u\bar{d}$-luminosity advantage of 8-TeV over 7-TeV running. For the cases mentioned in the preceding discussion, the 8-TeV $u\bar{d}$-luminosity multipliers are $\sqrt{\hat{s}} = 0.08\tev$: $1.15$; $0.175\tev$: $1.18$; $0.4\tev$: $1.22$; $1\tev$: $1.41$; and $2\tev$: $2.05$.  By $4\tev$, the advantage of $\sqrt{s} = 8\tev$ over $7\tev$ reaches an order of magnitude (see Figure~\ref{fig:udbar87log}).

\section{Light-quark--light-quark interactions}
\begin{figure}[p]
\centerline{\includegraphics[height=0.7\textheight]{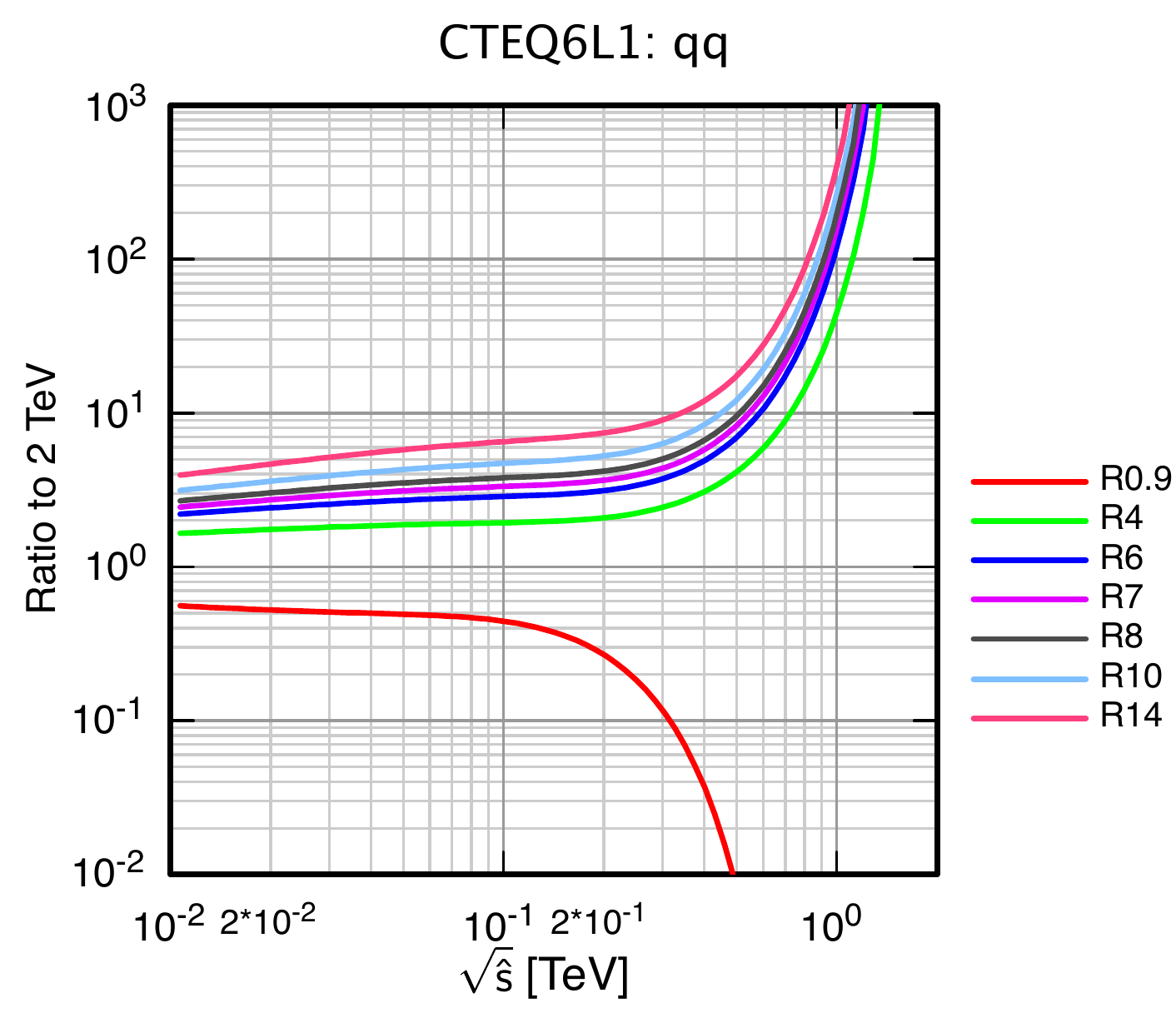}}
\caption{Comparison of parton luminosity for $qq$ interactions at specified energies with luminosity at $2\tev$.}
\label{fig:qqrat2}
\end{figure}

\begin{figure}[p]
\centerline{\includegraphics[height=0.7\textheight]{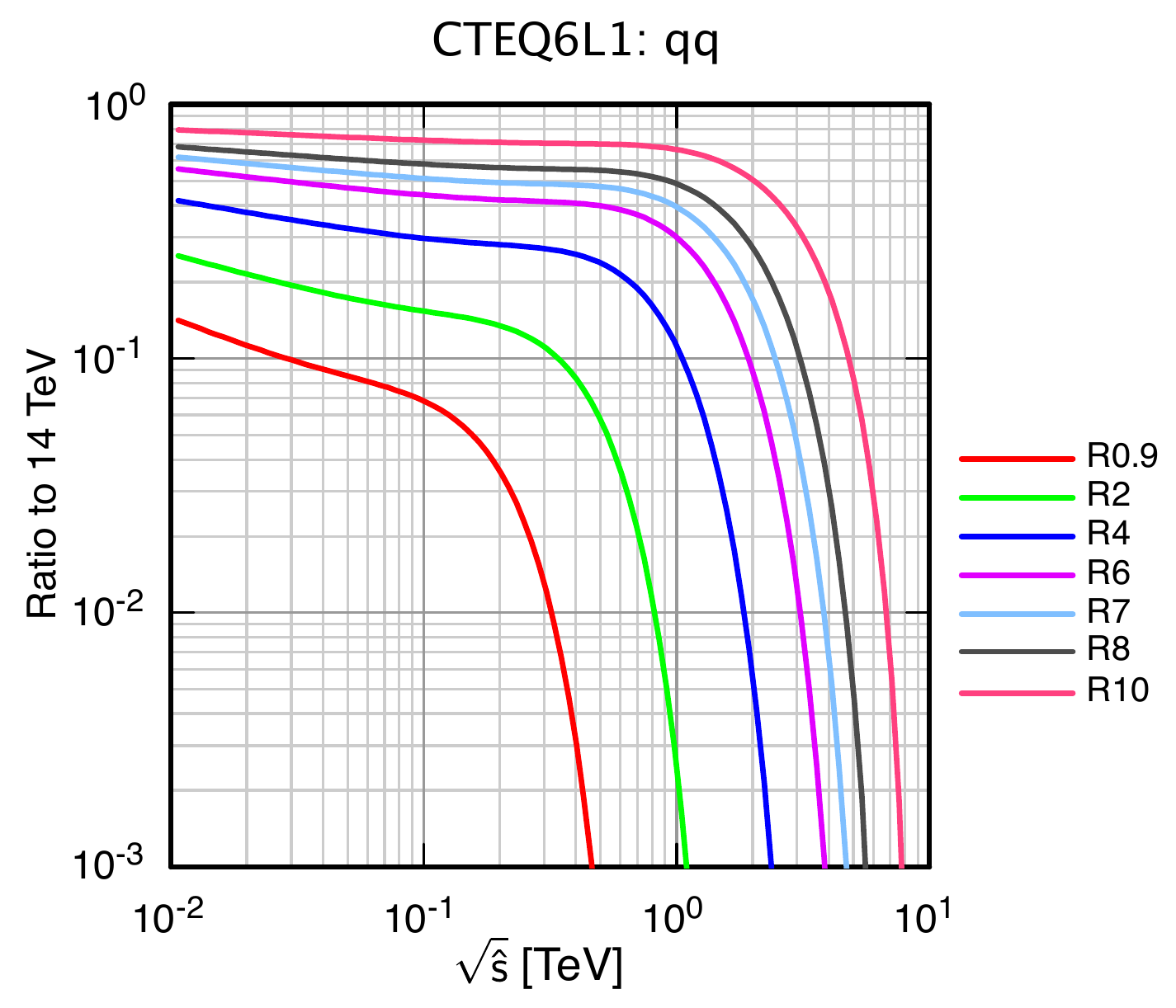}}
\caption{Comparison of parton luminosity for $qq$ interactions at specified energies with luminosity at $14\tev$.}
\label{fig:qqrat14}
\end{figure}

\begin{figure}[p]
\centerline{\includegraphics[height=0.7\textheight]{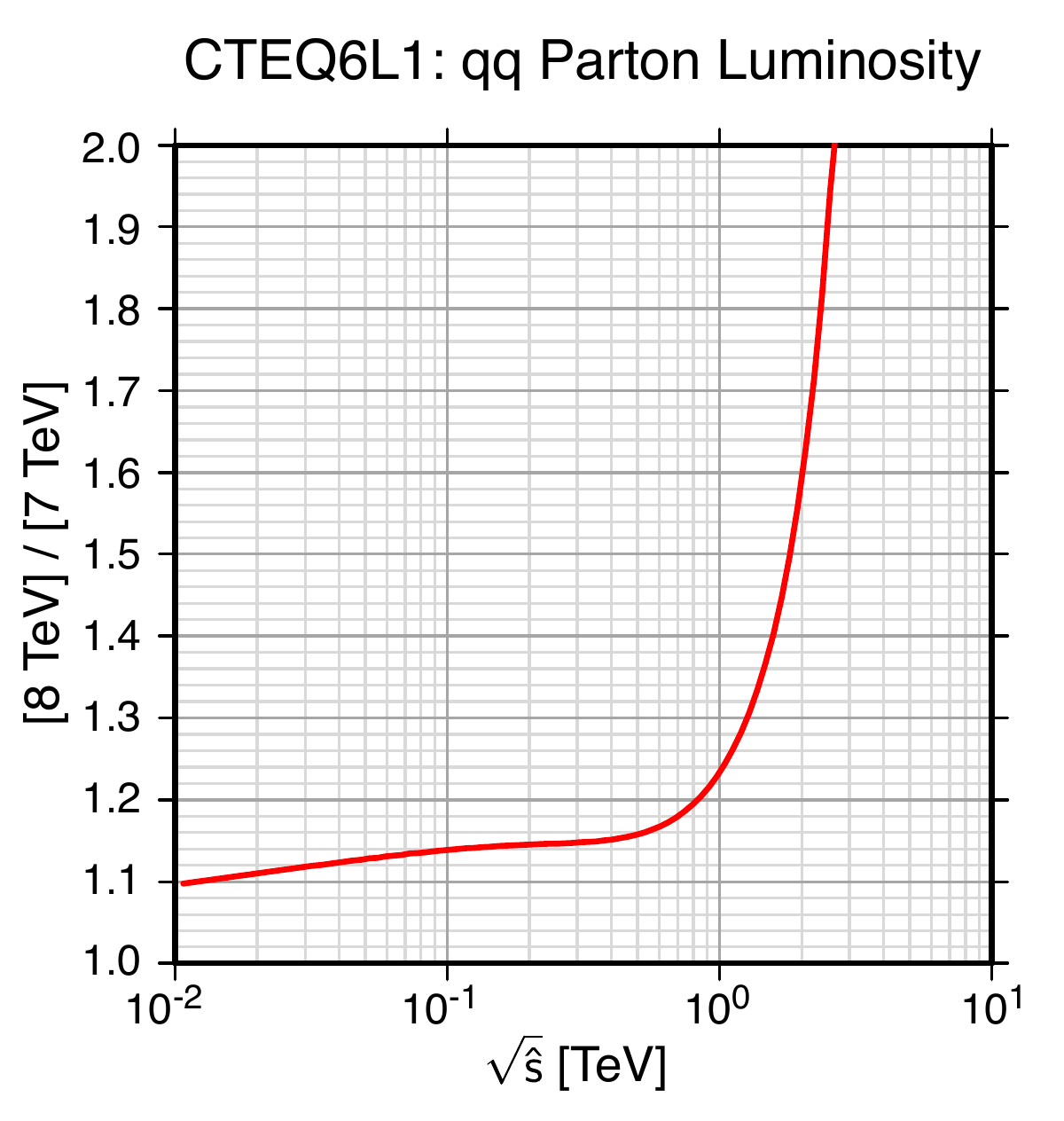}}
\caption{Ratio of parton luminosity for $qq$ interactions in $pp$ collisions at $\sqrt{s} = 8\tev$ to luminosity at $7\tev$.}
\label{fig:qq87}
\end{figure}

\begin{figure}[p]
\centerline{\includegraphics[height=0.7\textheight]{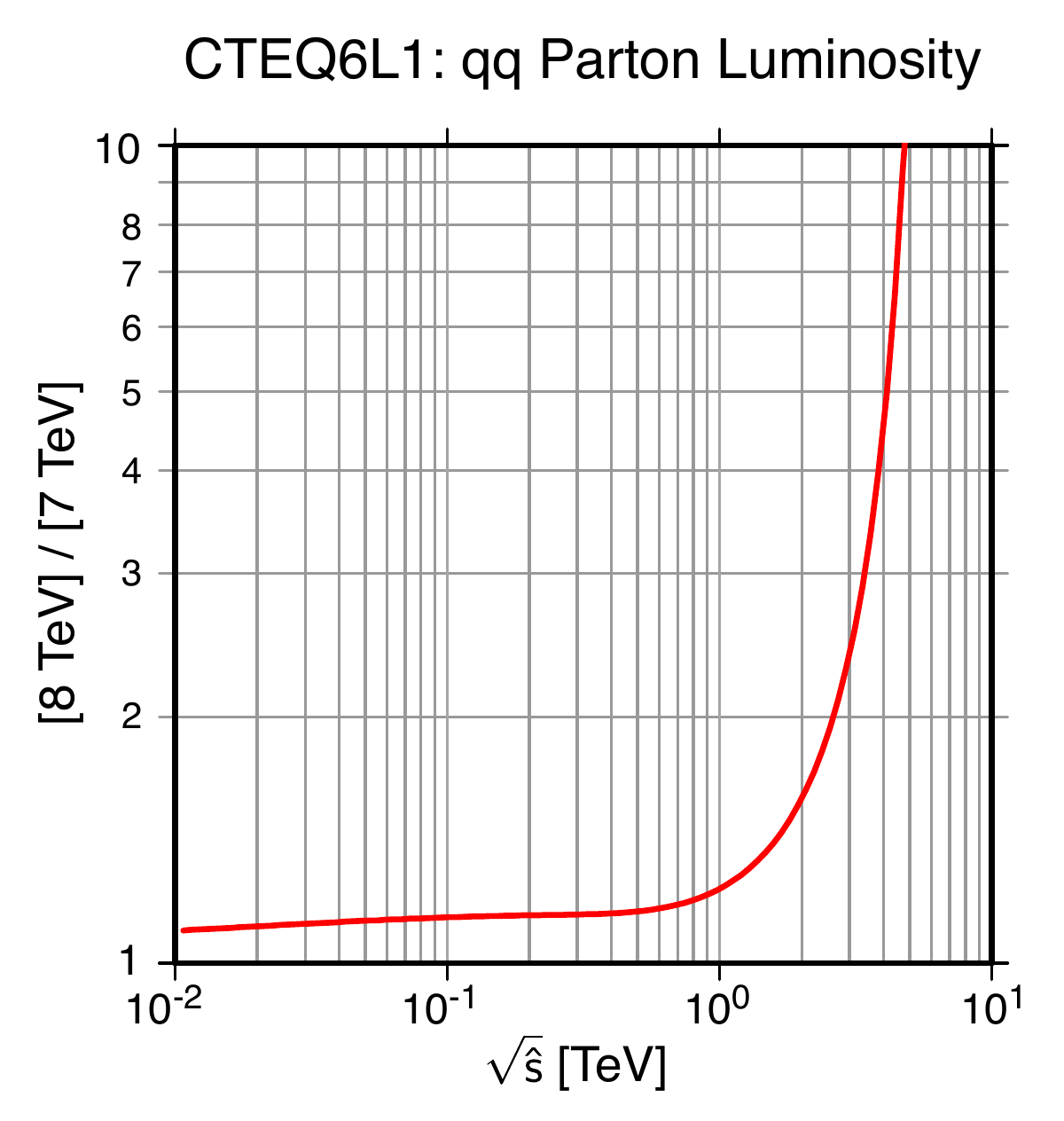}}
\caption{Ratio of parton luminosity for $qq$ interactions in $pp$ collisions at $\sqrt{s} = 8\tev$ to luminosity at $7\tev$ (logarithmic scale).}
\label{fig:qq87log}
\end{figure}

Ratios of parton luminosities for generic light-quark--light-quark interactions in $pp$ collisions at specified energies to the corresponding light-quark--light-antiquark interactions in $\bar{p}p$ collisions at the Tevatron are displayed in Figure~\ref{fig:qqrat2}. Ratios of the same luminosities  to the LHC at design energy appear in Figure~\ref{fig:qqrat14}. Such valence--valence interactions govern the production of hadron jets at very large values of $p_\perp$. Figure~\ref{fig:qqrat2} reveals that for jet production at $\sqrt{\hat{s}} \approx 1\tev$, $qq \to \hbox{two jets}$ will be enhanced at LHC energies $\sqrt{s} = (7, 8, 10, 14)\tev$ by factors of $(160, 195, 266, 400)$ over the rate for $q\bar{q} \to \hbox{two jets}$ at the Tevatron. At scales beyond those accessible at the Tevatron, a relevant question is how much rates are diminished in running below $\sqrt{s} = 14\tev$ (see Figure~\ref{fig:qqrat14}). For $\sqrt{\hat{s}} = 2\tev$, the rates at $\sqrt{s} = (7, 8, 10)\tev$ are $(0.17, 0.27, 0.51)\times$ those at design energy. The corresponding multipliers at $\sqrt{\hat{s}} = 4\tev$ are $(0.007, 0.031, 0.18)$. Figure~\ref{fig:qq87} exhibits the $qq$-luminosity advantage of 8-TeV running with respect to 7-TeV running. The 8-TeV luminosity multipliers are $\sqrt{\hat{s}} = 1\tev$: $1.23$; $2\tev$: $1.59$; and $4\tev$: $4.54$. For this valence-valence combination, the 
advantage of $\sqrt{s} = 8\tev$ over $7\tev$ reaches an order of magnitude at approximately $\sqrt{\hat{s}}= 5\tev$ (see Figure~\ref{fig:qq87log}).

\section{Gluon--light-quark interactions}
\begin{figure}[p]
\centerline{\includegraphics[height=0.7\textheight]{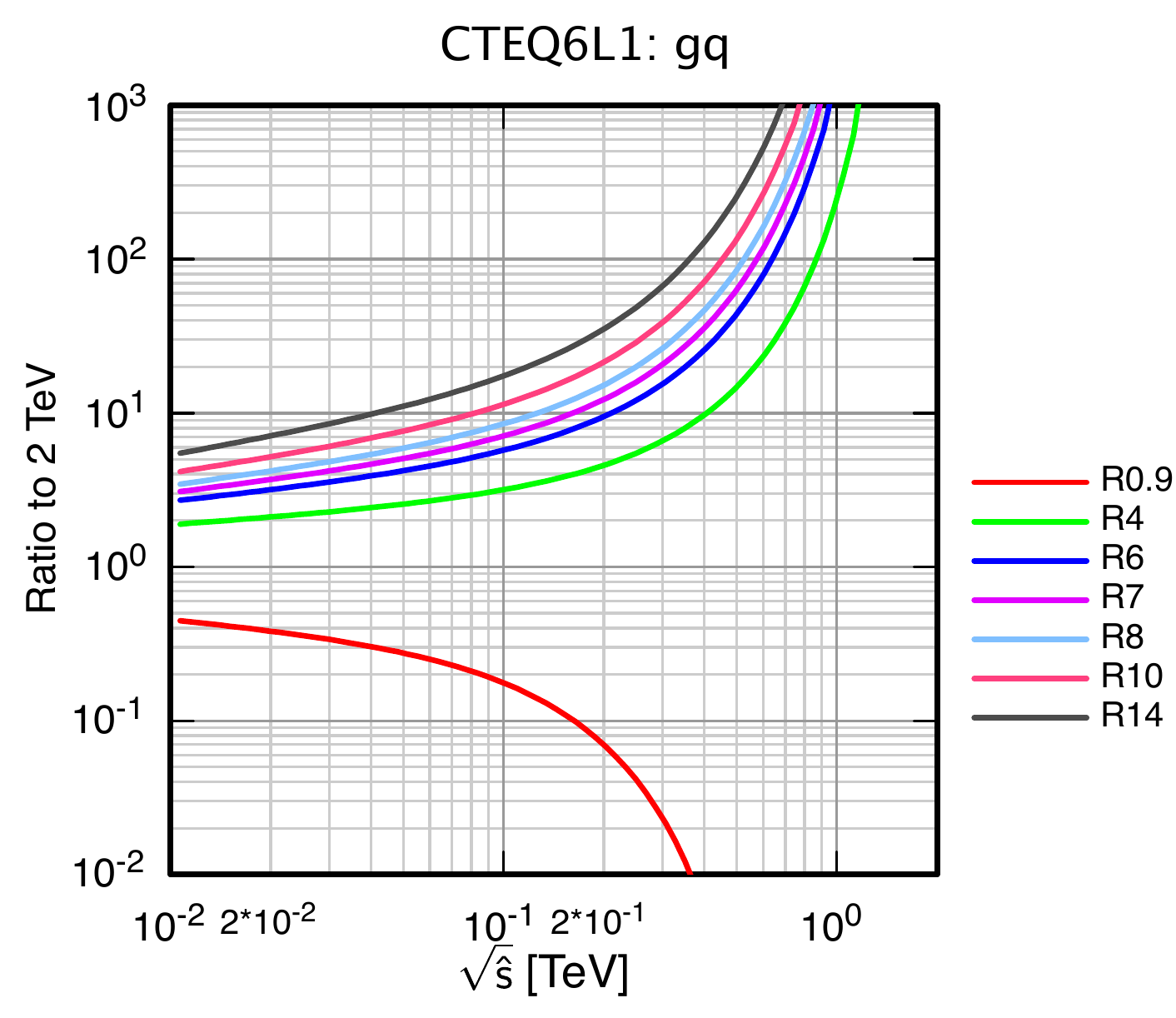}}
\caption{Comparison of parton luminosity for $gq$ interactions at specified energies with luminosity at $2\tev$.}
\label{fig:gqrat2}
\end{figure}

\begin{figure}[p]
\centerline{\includegraphics[height=0.7\textheight]{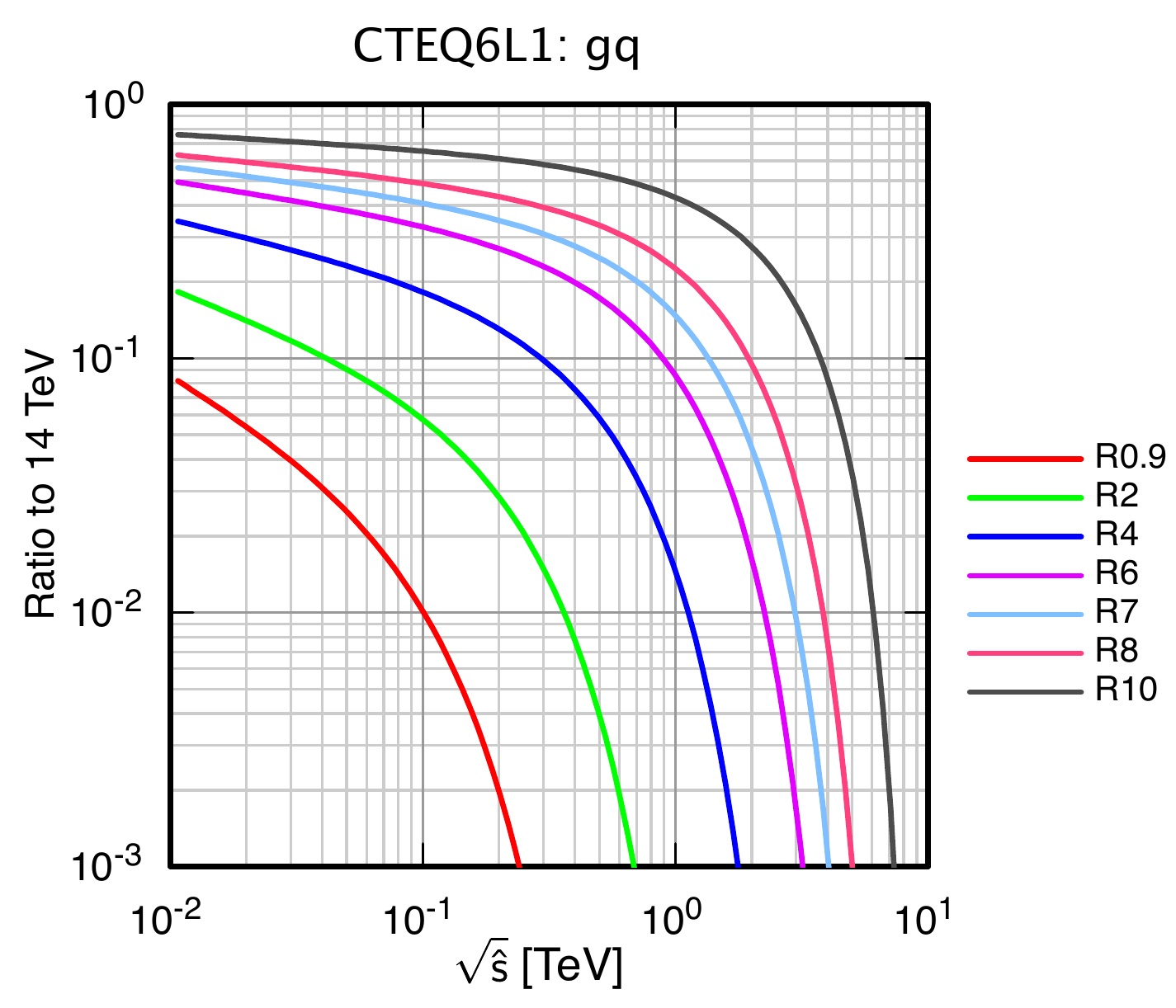}}
\caption{Comparison of parton luminosity for $gq$ interactions at specified energies with luminosity at $14\tev$.}
\label{fig:gqrat14}
\end{figure}

\begin{figure}[p]
\centerline{\includegraphics[height=0.7\textheight]{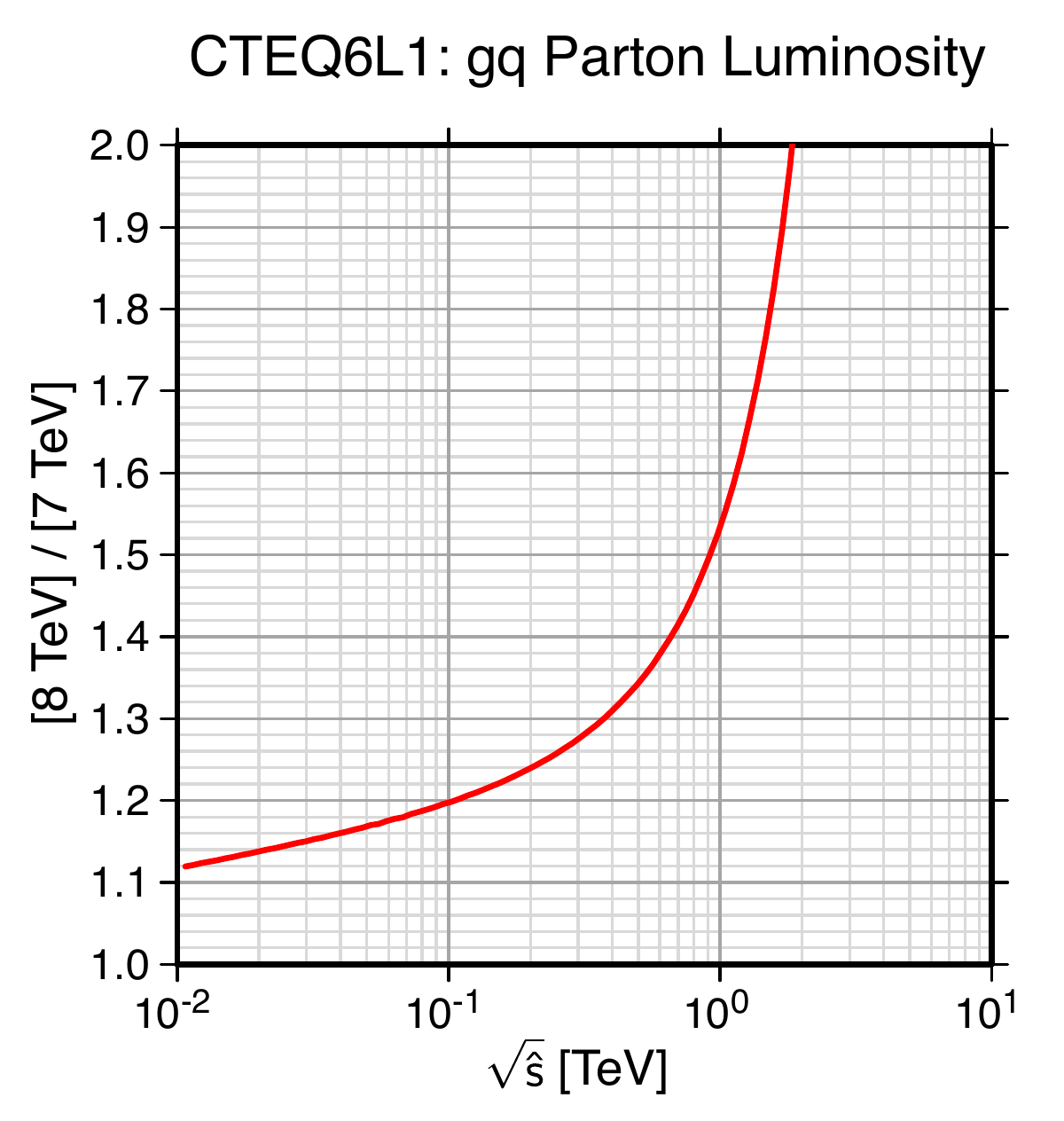}}
\caption{Ratio of parton luminosity for $gq$ interactions in $pp$ collisions at $\sqrt{s} = 8\tev$ to luminosity at $7\tev$.}
\label{fig:gq87}
\end{figure}

\begin{figure}[p]
\centerline{\includegraphics[height=0.7\textheight]{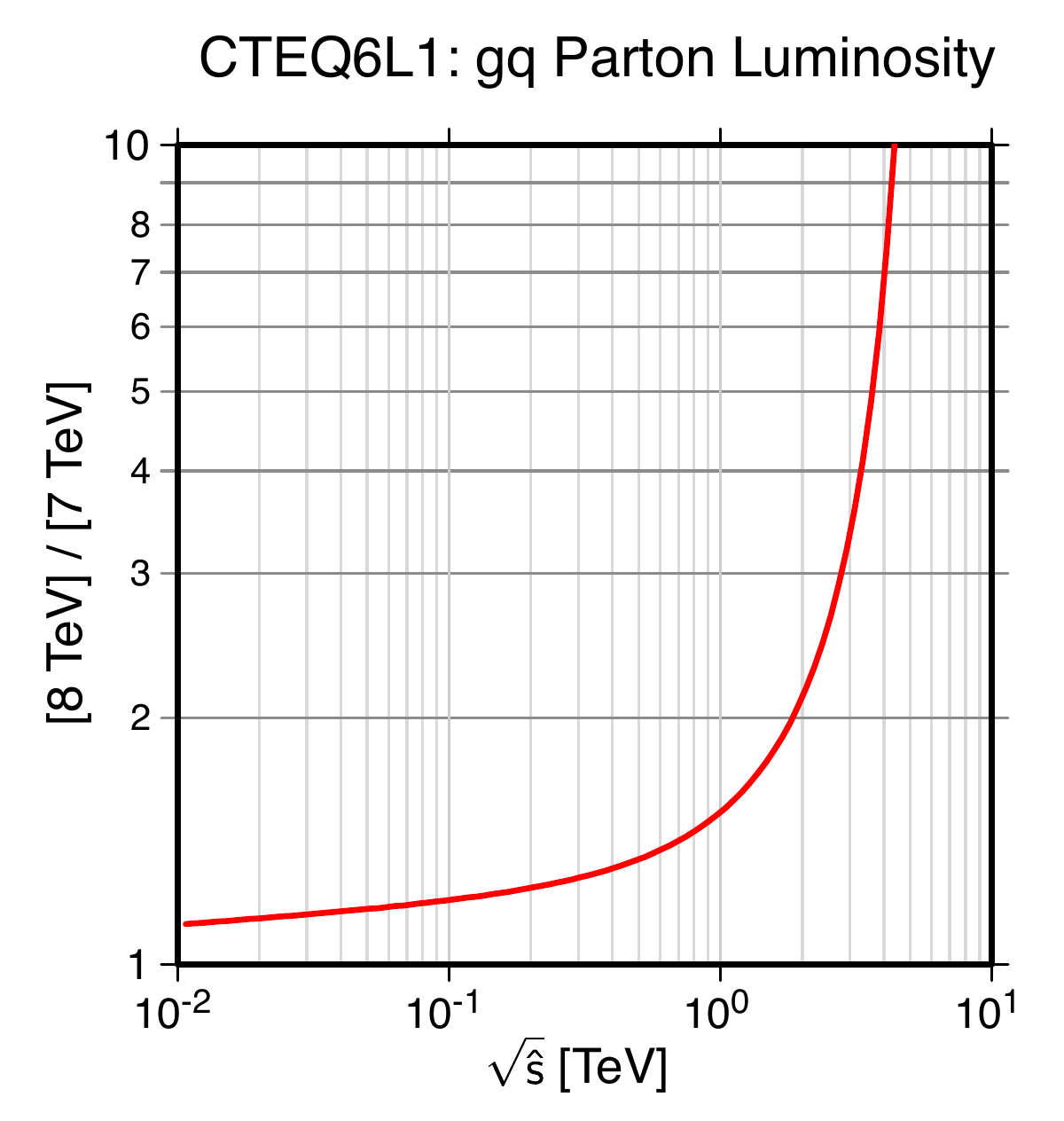}}
\caption{Ratio of parton luminosity for $gq$ interactions in $pp$ collisions at $\sqrt{s} = 8\tev$ to luminosity at $7\tev$ (logarithmic scale).}
\label{fig:gq87log}
\end{figure}

Collisions of gluons with light (mostly valence) quarks are important components of prompt-photon production and single--top-quark production, as well as jet production at intermediate values of transverse momentum. At $\sqrt{\hat{s}} = 300\gev$, typical for the Tevatron, $gq$ parton luminosities are higher than at the Tevatron by factors of $(21, 27, 39, 67)$ for $\sqrt{s} = (7, 8, 10, 14)\tev$, as shown in Figure~\ref{fig:gqrat2}. At the higher scale of $\sqrt{\hat{s}} = 1\tev$. the $gq$ luminosities at $\sqrt{s} = (7, 8, 10)\tev$ are smaller by factors of $(6.9, 4.4, 2.3)$ compared to the 14-TeV values (see Figure~\ref{fig:gqrat14}). The advantage of 8-TeV over 7-TeV running is a factor of $1.27$ at $\sqrt{\hat{s}} = 0.3\tev$ and $1.53$ at $\sqrt{\hat{s}} = 1\tev$ (see Figure~\ref{fig:gq87}).  Figure~\ref{fig:gq87log} shows that the relative advantage of $\sqrt{s} = 8\tev$ over $7\tev$ reaches an order of magnitude at $\sqrt{\hat{s}}\approx 4.4\tev$.

\section{Parton luminosity contours}
\begin{figure}[p]
\centerline{\includegraphics[height=0.7\textheight]{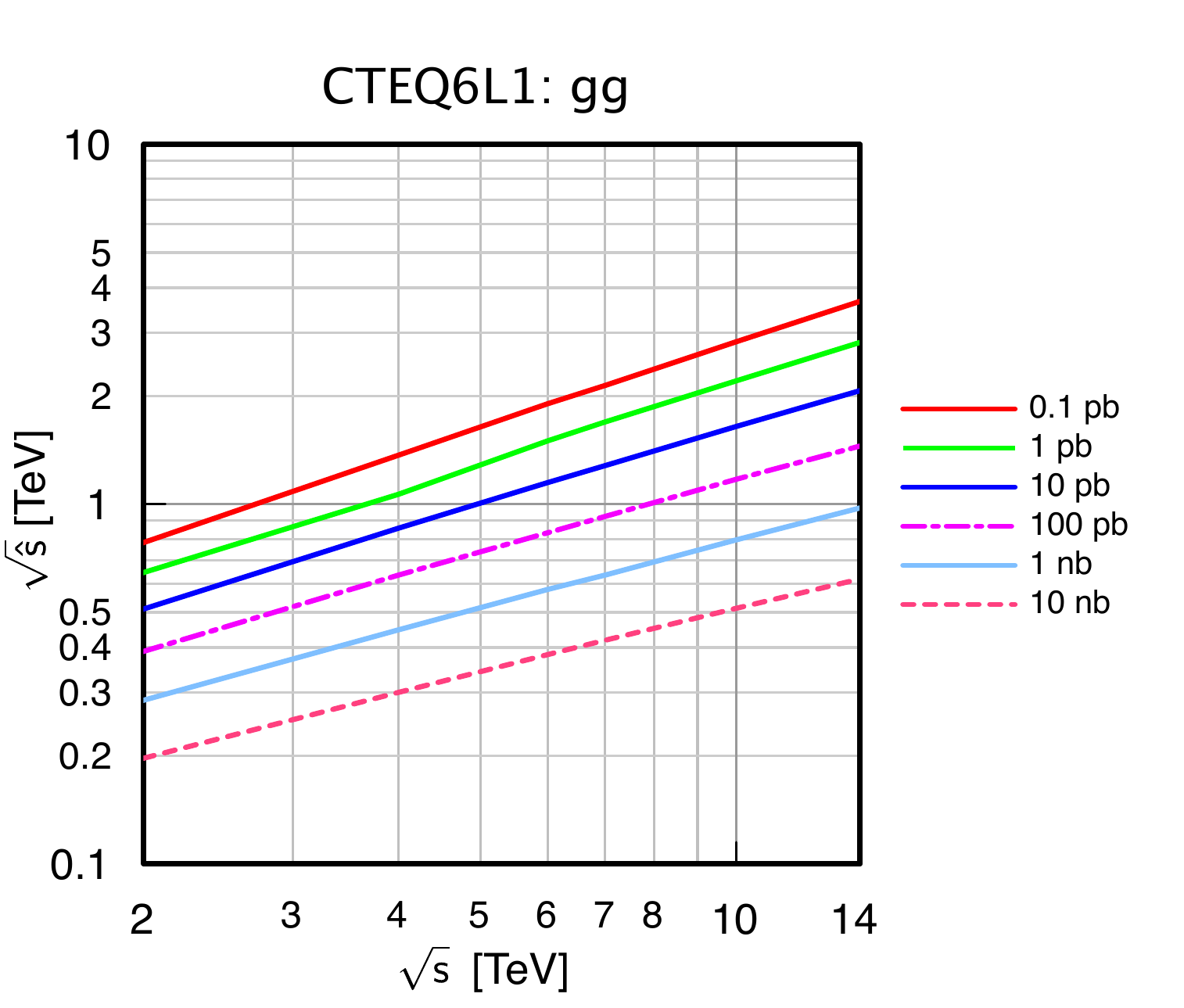}}
\caption{Contours of parton luminosity for $gg$ interactions in $p^\pm p$ collisions.}
\label{fig:ggcontours}
\end{figure}

\begin{figure}[p]
\centerline{\includegraphics[height=0.7\textheight]{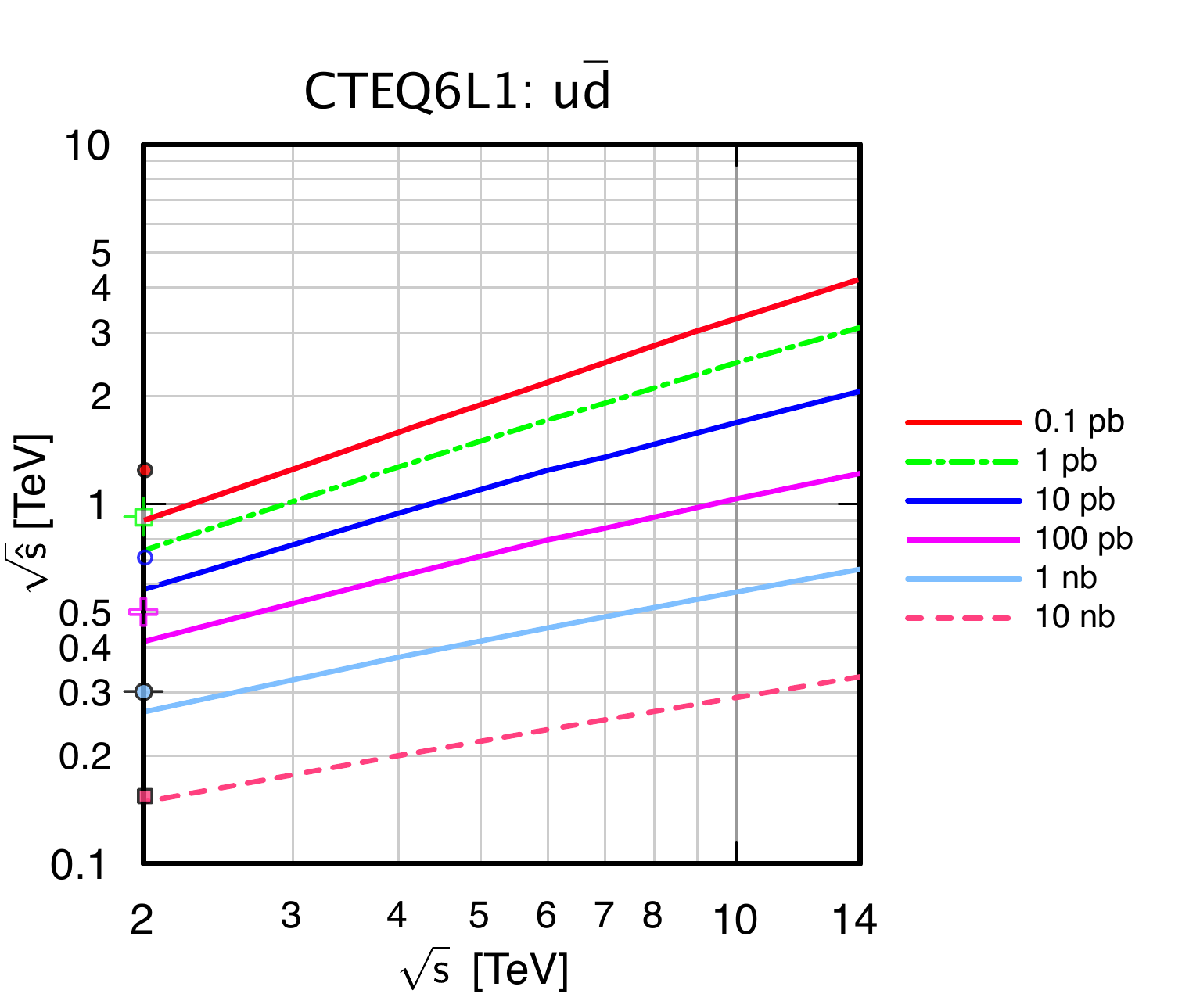}}
\caption{Contours of parton luminosity for $u\bar{d}$ interactions in $pp$ collisions. Values of $\sqrt{\hat{s}}$ corresponding to the stated values for $\bar{p}p$ collisions at the Tevatron are shown as points at $E = 2\tev$.}
\label{fig:udbarcontours}
\end{figure}

\begin{figure}[p]
\centerline{\includegraphics[height=0.7\textheight]{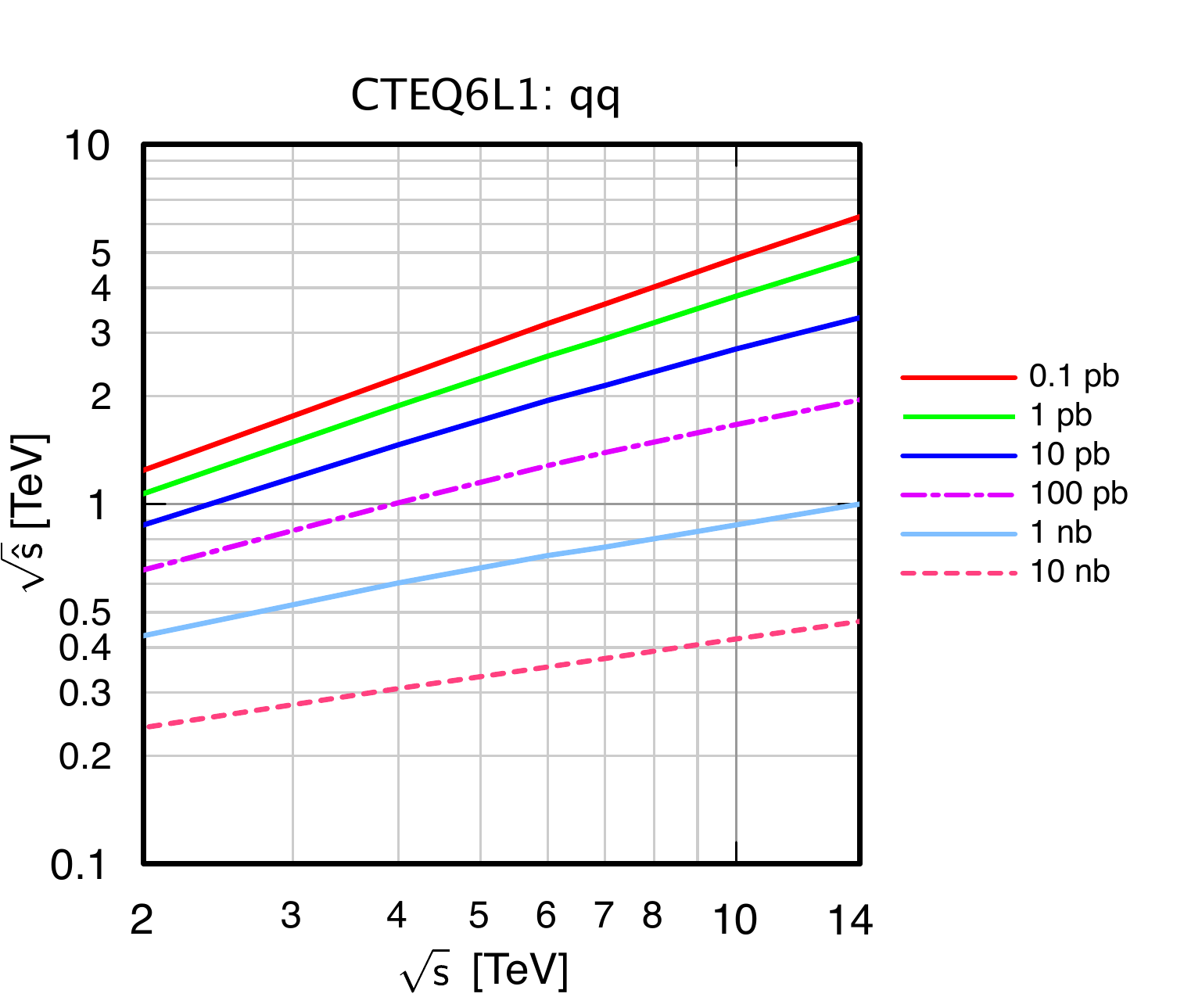}}
\caption{Contours of parton luminosity for $qq$ interactions in $pp$ collisions or $q\bar{q}$ interactions in $\bar{p}p$ collisions.}
\label{fig:qqcontours}
\end{figure}

\begin{figure}[p]
\centerline{\includegraphics[height=0.7\textheight]{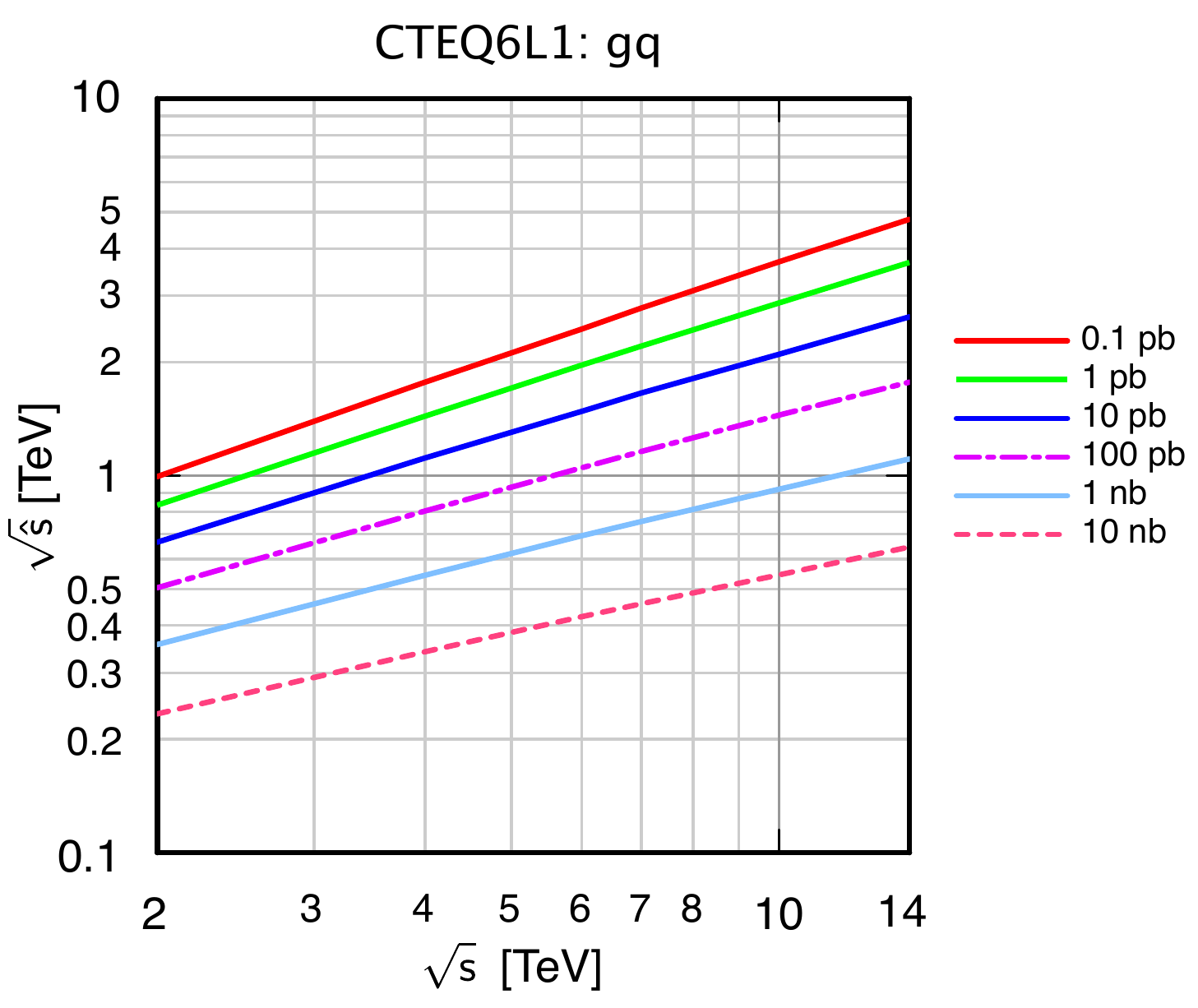}}
\caption{Contours of parton luminosity for $gq$ interactions ($\times{1}/{2}$) in $pp$ collisions and either  $gq$ or $g\bar{q}$ interactions in $\bar{p}p$ collisions.}
\label{fig:gqcontours}
\end{figure}

Contour plots showing at each hadron c.m.\ energy $\sqrt{s}$ the parton-parton energy $\sqrt{\hat{s}}$ that corresponds to a particular value of parton luminosity $(\tau/\hat{s})d\mathcal{L}/d\tau$ provide another tool for judging the effects of changes in beam energy or proton-proton luminosity. Plots for the four varieties of parton-parton collisions considered in this note are given in Figures~\ref{fig:ggcontours}--\ref{fig:gqcontours}, for proton-proton energies between $2$ and $14\tev$. The advantage of 2-TeV $\bar{p}p$ collisions over 2-TeV $pp$ collisions for $u\bar{d}$ interactions (at the same hadron luminosity) is indicated by the Tevatron points in Figure~\ref{fig:udbarcontours}.
The contour plots summarize a great deal of information, and will reward detailed study. As for the parton luminosity and ratio plots, contemplating the contour plots will be particularly informative to the user who brings a thorough understanding of signals and backgrounds at one or more beam energies.

\section{Final remarks}
Operational considerations will determine the beam energy for the 2011-2012 run of the Large Hadron Collider: safe, consistent operation that leads to high integrated luminosity must be the goal. Should it be practical to increase the beam energy from $3.5\tev$ to $4\tev$, the benefits to sensitivity are not inconsiderable. The benefits vary for different signals and backgrounds, so considering a representative sample of parton luminosities provides an efficient orientation.

Over the range of parton-parton collision energies from $\sqrt{\hat{s}} = 0.1\hbox{ to }1\tev$, the parton-luminosity enhancement for $\sqrt{s} = 8\tev$ compared with $7\tev$ ranges from about 15\% to 80\%, as recapitulated in Table~\ref{tab:sum}. The improvement is greatest for gluon-gluon collisions and least for quark-quark collisions. At higher scales, the effect is more pronounced: at $\sqrt{\hat{s}} = 4\tev$, the $8\tev:7\tev$ multiplier is an order of magnitude for $gg$ and $u\bar{d}$ interactions, $4.5$ for $qq$ collisions, and $6.9$ for $gq$ collisions. In the end, the product of hadron luminosity and parton luminosity is the figure of merit that influences physics performance. The $pp$ luminosity is likely to be higher (for fixed beam currents) at the higher beam energy. This would bring a double bonus.

The true promise of the LHC will be realized in high-luminosity running near the design energy. For the immediate future, accumulating several$\fb^{-1}$ at $\sqrt{s}=8\tev$ would begin a thorough exploration of the TeV scale and the origins of electroweak symmetry breaking.

\begin{table}[htdp]
\newcommand\T{\rule{0pt}{2.6ex}}
\newcommand\B{\rule[-1.2ex]{0pt}{0pt}}
\caption{Ratio of $\sqrt{s} = 8\tev$ to $\sqrt{s} = 7\tev$ parton luminosities.}
\begin{center}
\begin{tabular}{|c|c|c|c|}
\hline
Partons \T \B & $\sqrt{\hat{s}} = 0.1\tev$ & $\sqrt{\hat{s}} = 1\tev$ & $\sqrt{\hat{s}} = 4\tev$ \\
\hline
$gg$ \T & 1.28 & 1.81 & 9.72 \\
$u\bar{d}$ & 1.16 & 1.41 & 9.60 \\
$qq$ & 1.14 & 1.23 & 4.54 \\
$gq$ \B & 1.20 & 1.53 & 6.87 \\
\hline
\end{tabular}
\end{center}
\label{tab:sum}
\end{table}%

\footnotesize
\vspace*{-12pt}
\noindent Fermilab is operated by the Fermi Research Alliance under contract no.\  DE-AC02-07CH11359 with the
U.S.\ Department of Energy. I am grateful to Ignatios Antoniadis and the CERN Theory Group for warm hospitality.

\bibliographystyle{utphysrevA}

\bibliography{Luminosities}

\providecommand{\href}[2]{#2}\begingroup\raggedright\begin{thebibliography}{1}

\bibitem{Quigg:2009gg}
C.~Quigg, ``{LHC Physics Potential vs. Energy},''
  \href{http://arxiv.org/abs/0908.3660}{{\tt arXiv:0908.3660 [hep-ph]}}.
{FERMILAB-FN-0839-T}.

\bibitem{LHCjamboree}
``{LHC End-of-Year Jamboree, December 17, 2010},''
  \url{indico.cern.ch/conferenceDisplay.py?confId=113139}.

\bibitem{Alekhin:2010dd}
S.~Alekhin, J.~Blumlein, P.~Jimenez-Delgado, S.~Moch, and E.~Reya, ``{NNLO
  Benchmarks for Gauge and Higgs Boson Production at TeV Hadron Colliders},''
\href{http://arxiv.org/abs/1011.6259}{{\tt arXiv:1011.6259 [hep-ph]}}.

\bibitem{Eichten:1984eu}
E.~Eichten, I.~Hinchliffe, K.~D. Lane, and C.~Quigg, $\!$
  \href{http://dx.doi.org/10.1103/RevModPhys.56.579}{{\em Rev. Mod. Phys.} {\bf
  56}, 579--707 (1984)}.
\href{http://dx.doi.org/10.1103/RevModPhys.58.1065}{Addendum-\textit{ibid.}
  \textbf{58,} 1065 (1986).} See also R.~K.~Ellis, W.~J.~Stirling, and
  B.~R.~Webber,
  \href{http://theory.fnal.gov/people/ellis/BookFigs/Figs_jet/Figs7.html}
  {\textit{QCD \& Collider Physics}} (Cambridge University Press, Cambridge,
  1996) \S7.2; R.~K.~Ellis,
  \href{http://www.ippp.dur.ac.uk/Workshops/09/SUSSP65} {Scottish Universities
  Summer School in Physics 2009}; and plots on the
  \href{http://projects.hepforge.org/mstwpdf/plots/plots.html}{Martin-Stirling%
-Thorne-Watt parton distribution functions web site.}

\bibitem{Pumplin:2002vw}
{ {CTEQ Collaboration}}, J.~Pumplin {\em et al.}, $\!$
  \href{http://dx.doi.org/10.1088/1126-6708/2002/07/012}{{\em JHEP} {\bf 07},
  012 (2002)}
\href{http://arxiv.org/abs/hep-ph/0201195}{{\tt [arXiv:hep-ph/0201195]}}.

\bibitem{Stirling}
W.~J. Stirling, ``{Ratios of LHC parton luminosities},''
  \url{www.hep.phy.cam.ac.uk/~wjs/plots/lumi_LHC_789.eps}.

\end{thebibliography}\endgroup

 \clearpage
 
 \clearpage
 
\clearpage
 
\clearpage
 

\end{document}